\documentclass{aa}  

\AtBeginDocument{\nolinenumbers}
\let\linenumbers\nolinenumbers
\makeatother

\usepackage{graphicx}
\usepackage{txfonts}
\usepackage{lipsum}
\usepackage{subcaption}         
                                
\usepackage{lscape}             
                                
\usepackage{placeins}

\begin{document}

   \title{Insights into the structure and kinematics of a Milky Way-like galaxy}
   \nolinenumbers

   \author{E. Durán-Camacho\inst{1,2}\fnmsep\thanks{Corresponding author: eva.duran@iac.es}
        \and A. Duarte-Cabral\inst{3}
        }

   \institute{Instituto de Astrof\'{i}sica de Canarias, V\'{i}a L\'{a}ctea S/N, E-38205 La Laguna, Spain
   \and Departamento de Astrof\'{i}sica, Universidad de La Laguna, E-38206 La Laguna, Spain
   \and Cardiff Hub for Astrophysics Research and Technology (CHART), School of Physics and Astronomy, Cardiff University, Cardiff, CF24 3AA, UK}

   \date{Received 14 December 2025 / Accepted 26 February 2026}

\abstract
    {Understanding how the large-scale kinematics of the Milky Way (MW) shape the formation and evolution of the interstellar medium remains challenging from an observational perspective, and numerical models that can reproduce the observed structure and kinematics of the MW are much needed to be able to infer how the MW might work as a star formation engine.}
    {Our aim with this work is to use a numerical framework that is a close match to the observed large-scale distribution of stars and gas in the MW to isolate and understand the impact of galaxy-driven flows on the formation, agglomeration, and longevity of spiral patterns prior to the inclusion of chemistry, star formation, and feedback.}
    {We used an isothermal simulation of a MW-like galaxy found to closely match the longitude-velocity observational features of the MW by previous work that includes the coupled evolution of gas, stars, and dark matter under purely gravitational and hydrodynamical processes. 
    We characterised the morphology and kinematics of the stars and gas in the disc, quantified velocity residuals and their association with spiral features, and analysed the time evolution of individual spiral-ridge segments.}
    {Our results demonstrate that our model reproduces many of the observed structural and kinematic signatures of the MW, from the inner Galaxy to the solar neighbourhood, supporting its suitability as an analogue of the MW. The stellar spiral pattern in our model is relatively weak and shows a lower multiplicity relative to the sharper gaseous arms, offering an explanation for discrepancies in observational determinations of the number and location of MW spiral arms. Both the gas and stellar spiral arms are highly segmented, without a single coherent spiral pattern that would be expected from a grand-design type of galaxy. We find strong radial motions linked to the non-circular motions driven by the presence of a bar and that extend well into the disc. The gas radial and tangential velocity residuals can be as strong as $30$--$50$ km\,s$^{-1}$,  with alternating patterns of converging and diverging flows (promoting the growth and dissolution of spiral arm segments), and they evolve over short timescales of $\sim$10--20 Myr. This transient, dynamically driven nature of spiral structures could explain the observed low contrast in cloud properties and star formation occurring inside versus outside spiral arms in the MW.}
    {}

   \keywords{Galaxy: structure --
                galaxies: spiral --
                ISM: structure --
                methods: numerical
               }

   \maketitle

\section{Introduction}
\label{sec:intro}
Understanding the formation of stars is fundamental to comprehending galactic evolution, the interstellar medium (ISM), and the overall evolution of the Universe. Star formation processes influence the chemical enrichment of the ISM and distribution of energy within galaxies. As a consequence, there is a substantial effort to determine whether star formation is significantly influenced by the large-scale structure and the dynamics of galaxies or whether it depends solely on the amount of dense gas available. The Milky Way (MW) is a prime laboratory for resolved studies of star formation within molecular clouds, but our position within the Galactic disc limits our ability to link the star formation properties of clouds to their larger-scale Galactic context -- with large distance uncertainties to individual clouds making it incredibly hard to pin-point their exact location within a highly uncertain Galactic structure. Thus, despite extensive knowledge of star formation within individual molecular clouds, the extent to which the MW's large-scale structure affects or regulates these processes remains unclear. Numerical models can aid in understanding how the MW might act as a star-formation engine, but most MW-type galaxy models do not necessarily match the observed structure and kinematics of our Galaxy, thus limiting our ability to draw firm conclusions regarding the interpretation of trends in our own Galaxy. With this work, we aim to provide a framework to further our understanding of the MW via a model that attempts to reproduce our current combined observational knowledge of the MW structure and dynamics.

\label{sec:intro-inner}

The Milky Way has been known to be a barred galaxy since the early 1960s, initially identified via gas velocities \citep{de1964comparison} and subsequently confirmed via H$_{\textsc{I}}$ and CO emission \citep{binney1991understanding}, photometry in the near-infrared \citep{blitz1991direct,weiland1994}, and star counts \citep{weinberg1992detection,stanek1994, Dwek1995}. Studies have suggested that the bar extends past the central bulge, with a semi-major axis between 3.7-4 kpc \citep{hammersley1994infrared}, though debate exists regarding the structure of the bulge-bar system. Some studies suggested a triaxial bulge \citep{lopezcorredoira2005}, while others propose a double-barred structure \citep{benjamin2005, cabrera2007tracing}. Recent \textit{Gaia} data have confirmed a bar structure with a length of $\sim$4 kpc, bar orientation of $20^{\circ}$, and pattern speeds around 38-41 km s$^{-1}$ kpc$^{-1}$ \citep{Bovy2019, queiroz2021milky, drimmel2023}.

Numerical models have supported these findings, with \cite{portail2017dynamical} estimating a pattern speed of $\Omega_{\mathrm{p}}= 40$ km s$^{-1}$ kpc$^{-1}$. Other simulations, such as those by \cite{sormani2022stellar}, have refined this estimate to $\Omega_{\mathrm{p}} \sim 37.5$ km s$^{-1}$ kpc$^{-1}$, in line with observations of the Milky Way's boxy/peanut-shaped bulge structure \citep{clark2019tracing, clarke2022, li2022}. These studies collectively reinforce the characterisation of the Milky Way's bar as a key component of its overall structure.

In terms of kinematics, gas in a non-axisymmetric potential, such as that of a barred galaxy, follows two different types of closed orbits near the Galactic centre: $x_{1}$-type that represents elongated orbits parallel to the bar major-axis and $x_{2}$-type orbits that are perpendicular to the major-axis of the bar \citep{binney1991understanding}. These orbits result from the bar dynamics and are dependent on the bulge size and mass \citep[see e.g.][]{bureau1999nature}. Indeed, the existence, size, and extent of $x_{2}$ orbits are known to be influenced by the location of the inner Lindblad resonance (ILR) \citep[e.g.][]{Contopoulus1989,athan92a}. \cite{Sparke1987} argued that this family of orbits is nearly empty in the absence of gas, but when gas is present, the bar drives the gas inwards, and it settles into $x_{2}$ orbits \citep{binney1991understanding}. The conventional understanding of bar orbital structures claims that the primary types of bar orbits are quasi-periodic or regular orbits originating from stable $x_{1}$ and $x_{2}$ orbits \citep[e.g.][]{Contopoulus1980,Athanassoula1983}. \cite{Abbot2017} examined the orbits primarily responsible for a boxy/peanut or X-shaped bulge shape (as found in the centre of the MW) and found that between $19-23 \%$ of the bar's mass is linked to the X-shape, which is significantly influenced by various bar orbit families, including non-resonant box, `banana', fish/pretzel', and `brezel' orbits. No single family accounts for all observed features, but since \cite{Valluri2016} showed that box orbits are predominant in bars and highly adaptable due to their unique frequencies, these are proposed as the most significant in shaping the X-bulge.

\label{sec:intro-disc}

Turning to the Galactic disc, the nature and number of spiral arms in galaxies such as the MW are subject of intense study and debate. The formation and persistence of these spiral structures, as reviewed by \cite{dobbs2014dawes}, are influenced by a variety of dynamical processes. Key theories include the density wave theory, which proposes that spiral arms are density enhancements rotating through the disc, and gravitational instabilities contribute to the spontaneous emergence of spiral patterns. Additionally, tidal interactions with other galaxies can induce spiral structures, while the interplay between gas and stellar dynamics further shapes their appearance. And finally, N-body simulations suggest that there can be a more transient nature to the spiral structure of galaxies, where spiral arms are more dynamic short-lived features organised by gravitational forces and recurrent patterns regulated by the heating and cooling processes within the disc \citep[e.g.][]{Sellwood1984,baba2015dynamics,sellwood2019spiral,pettitt2020spiral}. For this type of spiral structure, spiral arms are not single well-defined entities. Instead, they are constructed as a combination of superimposed arm-segments, complicating their detection and identification. 
This diversity in formation mechanisms leads to a variety of spiral arm structures observed across the galaxy population, from grand design spirals to flocculent and multi-armed galaxies, underscoring the complexity in determining a universal model for spiral arm formation and consequently the role that spiral arms might have on star formation. 

The core of the debate about the number of spiral arms in the Milky Way lies in determining whether our Galaxy is characterised by a predominantly two-arm spiral structure, as some studies propose \citep{drimmel2000evidence}, or if it instead features a more complex configuration with four or more arms \citep[e.g.][]{churchwell2009}. The discrepancies on the observed position and number of arms can be attributed to several factors. For one, different observational methods and wavelengths highlight various components of the arms, such as star-forming regions versus overall mass distribution. 
Optical observations often emphasise the locations of young, massive stars due to their brightness, which can suggest a simpler two-arm structure. However, optical data can capture a broad range of stellar populations \citep{drimmel2001}. In contrast, infrared observations, such as those from the Spitzer GLIMPSE survey, are sensitive to star-forming regions and older and cooler stars, as they are particularly good at penetrating dust. These findings have revealed a more complex structure with four or more arms, including several fainter arms and spurs \citep{benjamin2005,churchwell2009}. Radio observations of neutral hydrogen, which map the overall gas mass more comprehensively, also support a multi-armed configuration with four or more arms \citep{levine2006,Hong2022}.

Observations of the molecular gas distribution in our Galaxy, also suggest the existence of multiple arms \citep[e.g.][]{dame2001milky}, although their nature is still unclear. For instance, results from the observations of the first Galactic quadrant \citep[e.g.][]{ROMAN-DUVAL2010,rigby2016chimps,rigby2019} reveal enhancements in the surface density of molecular gas along some arms, and lower linewidths in the inter-arms, suggesting that molecular clouds are dynamically forming in spiral arms and may be disrupted in inter-arm spaces, which is more in line with grand-design type of pattern. That, however, is in contrast with the results from the SEDIGISM survey towards the fourth quadrant \citep[e.g.][]{duarte2021sedigism,colombo2022sedigism}, where no significant contrast was found between the properties of molecular clouds in spiral arms and inter-arm regions, pointing towards a more transient nature of the arms. This disparity, however, could be partly due to the first quadrant being more affected by the bar dynamics and inducing stronger trends.
 
It is worth noting, however, that most observational studies (except for those that use parallax measurements), have very high uncertainties in placing objects in their 3D position within the Galaxy. Particularly for the gas, observations use the observed line-of-sight velocities to infer their kinematic distances (by assuming an idealised rotation curve, which can be heavily distorted by any proper-motion of the gas). Hence, in order to better understand the link between the observed kinematics of the gas, and its inferred 3D structure, \cite{Ramon-fox2018} performed high-resolution smoothed particle hydrodynamics (SPH) simulations of gas dynamics within spiral arms of a Milky Way-type model with a fixed four-arm spiral potential (i.e. mimicking a density wave), without a bar.
Their findings indicate that the spiral arms induce net radial motions of $\Delta \mathrm{V}_{R}$ $\sim -9$\,km\,s$^{-1}$ and azimuthal motions of $\Delta \mathrm{V}_{\Phi} \sim 6$ km\,s$^{-1}$ slower than the rotation curve. Such dynamics lead to systematic errors in kinematic distance estimates by $\sim 1$ kpc. Consequently, observers mapping cloud positions based on these distances might encounter distortions and systematic offsets from the actual structures of up to $\pm 2$\,kpc. These results are, of course, valid for this type of imposed potential that mimics a density wave, and this could become more complex if indeed the spiral arms are more dynamic in nature, and if the bar potential affects the dynamics of the disc further. Thus, validating simulation outputs against empirical data from observations becomes crucial for refining our understanding of Galactic structure.

Missions such as \textit{Gaia} and surveys such as APOGEE \citep[e.g.][]{katz2018gaia,Antoja2021,Majewski2017,Bland-Hawthorn2016,Minniti2010} have provided unprecedented precision in the measurements of positions, velocities, and chemical compositions of stars across the Galaxy. In particular, \textit{Gaia} Data Release 3 \citep[DR3;][]{GaiaDR32021} has enabled detailed kinematic maps within a few kiloparsecs from the Sun. For example, \citet{khanna2023measuring} revealed intricate streaming-motion patterns in the Galactic disc that likely encode the imprint of non-axisymmetric structures. \cite{drimmel2023} further explored the disc of the Milky Way and highlighted the asymmetries in the velocity fields and spatial distributions of stars, attributing these features to non-uniform mass distributions and ongoing dynamical processes influenced by the Galactic bar and spiral arms, which are more evident in the outer regions of the Galaxy. At the same time, \textit{Gaia} has accelerated efforts to map the Milky Way spiral structure using young stellar tracers, although the inferred morphology remains uncertain and tracer-dependent due to extinction, selection effects, and our location within the disc. A recent overview of these developments in Galactic dynamics and their interpretation in the \textit{Gaia} era has been provided by \citet{HUNT2025}. Complementary spiral-arm constraints come from radio maser parallaxes \citep[e.g.][]{reid19,Marshall2025} and standard-candle populations, such as classical Cepheids. In particular, \citet{Drimmel2025} have presented a recent parametric spiral-arm model based on Cepheids with mid-infrared distance estimates, extending spiral mapping to distances where optical tracers are limited \citep[see also][]{xu2023review}. Overall, the data from \textit{Gaia} DR3+ have revealed a more turbulent and dynamically active Milky Way than previously thought, challenging existing models to account for the observed complexity.

A complementary avenue for studying MW-type galaxies involves fully cosmological simulations, which follow the assembly history, merger events, and large-scale environment of haloes similar to that of the MW \citep[e.g.][]{grand2017auriga, Wang2015, Buck2020, Schaye2015, Stinson2012, Serrato2023}. These models are invaluable to understanding the statistical diversity of MW-type galaxies and to assessing how cosmological accretion and mergers shape galactic evolution. However, they are not suitable for the purpose of helping interpret the specific properties of the MW, as that requires a fine-tuned MW-analogue that reproduces the detailed structural and kinematic properties of the MW itself. Furthermore, the high computational cost of tracking the evolution of a cosmological system makes it prohibitive to re-simulate the same system at a substantially higher resolution or under varied physical prescriptions. For these reasons, and as discussed in \citet{DuranCamacho2024}, we adopted a controlled isolated-galaxy framework that allows the initial conditions to be tuned such that the resulting system resembles the present-day MW as closely as possible. This approach provides the flexibility and resolution needed to build a practical dynamical model of the MW suitable for studying the origin of its present-day structures and kinematic patterns.

In \citet{DuranCamacho2024}, we explored a large suite of MW-type models with live gravitational potentials, which enabled a dynamic and self-consistent evolution of the gas and stars, unlike models with imposed external fixed potentials \citep[e.g.][]{dobbs2013exciting,ridley2017nuclear,smith2020cloud}. In that work, we identified the model that was the closest reproduction of the MW structures and terminal velocities as seen in longitude-velocity space (\textit{lv} space). In this follow-up study, we further test whether this model provides a robust dynamical description of the MW by benchmarking it against additional observational constraints and against studies modelling specific Galactic components \citep[e.g.][]{sormani2022stellar,portail2017dynamical,ridley2017nuclear}. We also investigate the role of purely galaxy-driven kinematics in shaping the agglomeration, evolution, and disruption of spiral patterns in the model (thus prior to the inclusion of chemistry, star formation, and feedback, which will be the subject of future work). We use these results to help deepen our understanding of the interplay between stellar and gaseous motions across the disc, bulge, and bar, and we place the results in the context of recent \textit{Gaia} DR3 findings and ongoing debates on the nature of the MW spiral structure.

The remainder of this paper is organised as follows: In Section~\ref{Section: Model Overview}, we summarise the hydrodynamical simulation used in this work. In Section~\ref{section: inner Galaxy} we investigate  our model's stellar structure and kinematics in the inner galaxy and benchmark our results against observationally motivated analytical models.  In Section \ref{sec:Galactic_disk}, we investigate the properties of the galactic disc in our model, in particular the spiral structure (Section~\ref{sec: spiral pattern}) and the distribution of gas and stars around the solar neighbourhood (Section~\ref{Subsec: stellar disc kinematics}). In Section~\ref{Sec:kinematics_large_structures} we investigate the link between the large-scale kinematic patterns and the formation of high-density structures, in particular by analysing the gas kinematics along different cross-sections of the disc (Section~\ref{Subsec: gas disc kinematics}), and by following the creation and destruction of spiral arm segments over time (Section~\ref{Subsec:time_evolution_spiral_arm}). Finally, we discuss and summarise our findings in Section~\ref{Sec: Conclusions}.

\section{Model overview}
\label{Section: Model Overview}

The simulation analysed in this study was introduced in 
\citet[][]{DuranCamacho2024} as their Model 4. For conciseness, here we only provide a brief description of the numerical setup, and readers are directed to \citet[][]{DuranCamacho2024} for a more in-depth description of all the parameters. The model was generated using the {\sc arepo} numerical code \citep{springel2005cosmological}, using a live potential of dark matter particles (following a spherical Hernquist profile), star particles (distributed in a disc with an exponential surface density profile, and a spherical stellar bulge following an Hernquist profile), and gas cells (distributed in a disc with an exponential surface density profile that follows that of the stars). These initial conditions were generated using the {\sc makenewdisc} code \citep{springel2005modelling}. For the specific parameters that define each of these distributions see \citet[][]{DuranCamacho2024}. From a suite of 15 models with varying initial stellar mass distributions, \citet[][]{DuranCamacho2024} identified Model 4 (at a time of 2.6 Gyr) as the one that most accurately matched the main observed structure of the Milky Way, as seen in \textit{lv} space, using $^{12}$CO and \ion{H}{I} data obtained from \cite{dame2001milky} and \cite{bekhti2016hi4pi}. That model has a total stellar mass of $4.25 \times 10^{10}$\,M$_{\odot}$ (of which 10\% is in the bulge), a dark matter mass of $8.97 \times 10^{11}$\,M$_{\odot}$, and a total gas mass in the disc of $8.6 \times 10^{9}$\,M$_{\odot}$.

The model was run with isothermal equations of state with a sound speed set to $c_{s} = 10$\,km\,s$^{-1}$, and without including gas self-gravity. The gas resolution within our simulation adheres to a refinement criterion based on the mass and volume of gas cells, targeting a gas mass of $1000$\,M${\odot}$ per cell. The model has minimum and maximum cell volumes of $27$\,pc$^{3}$ (equivalent to a cube of $3$\,pc side) and $1.25 \times 10^{8}$\,pc$^{3}$ (equivalent to a cube of $500$\,pc side), respectively. The stellar particles have a mass resolution of $8000$\,M$_{\odot}$ and a softening length of $50$\,pc, while the dark matter particles have a mass of $9 \times 10^{5}$\,M$_{\odot}$ and a softening length of $300$\,pc. The simulation was conducted within a cubic box of $100$\,kpc on each side, employing periodic boundary conditions\footnote{ This box is only required for the hydrodynamical part of the code, i.e. the gas cells. All other particles (stars and dark matter halo) are effectively free to take up any position without any bounds. In practice, this means the dark matter halo extends well beyond the 100\,kpc box.}. The gas within the galactic disc is mostly distributed within the central $\sim$30\,kpc of the box, thus far enough from the box boundaries such that this boundary condition has no impact on the evolution of the system.

\section{The inner galaxy}
\label{section: inner Galaxy}

Building on the work from \citet{DuranCamacho2024}, here we investigate in more detail  the inner galaxy structure of our simulation, in order to test where and how it is able to capture some of the properties of the inner MW. This is an important benchmark required to inform potential other applications of this model as a MW analogue. In this section, we focus our analysis on the structure and kinematics of both stars and gas, and we compare them with observationally inferred results.

\subsection{Stellar distribution}
\label{subsection: inner Galaxy - morphology}

\begin{figure*}
	\includegraphics[height=7.5cm]{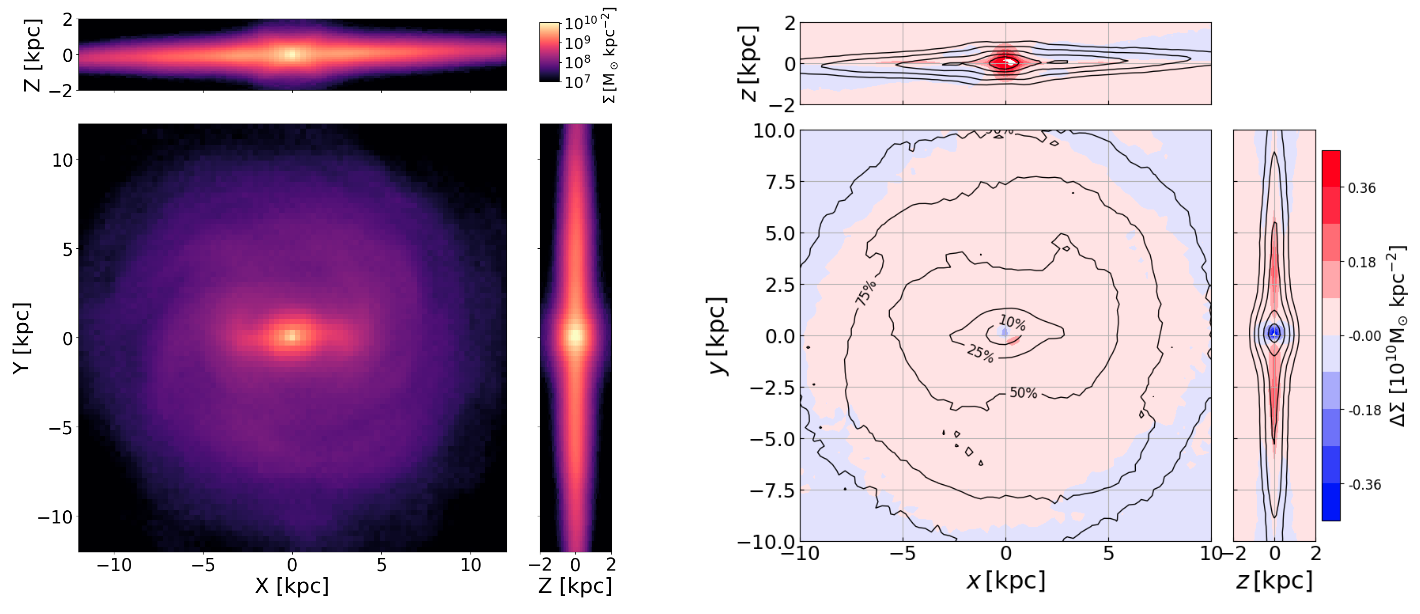}
    \hfill
    \caption{Left: Column density maps of the stellar distribution within our model, projected in the $xy$, $xz$ and $yz$ planes.  Right: Residual column density maps of the stellar disc distribution within our model compared to S22. The contours represent the cumulative fractions of the total stellar mass our model: $10\%, 25\%, 50\%, 75\%,$ and $90\%$. This figure uses colour-coding, where red indicates a lack and blue denotes an excess of mass in our model, compared to the S22 analytic distribution. }
    \label{fig:stellar disc}
\end{figure*}

\begin{figure*}
	\includegraphics[width=\textwidth]{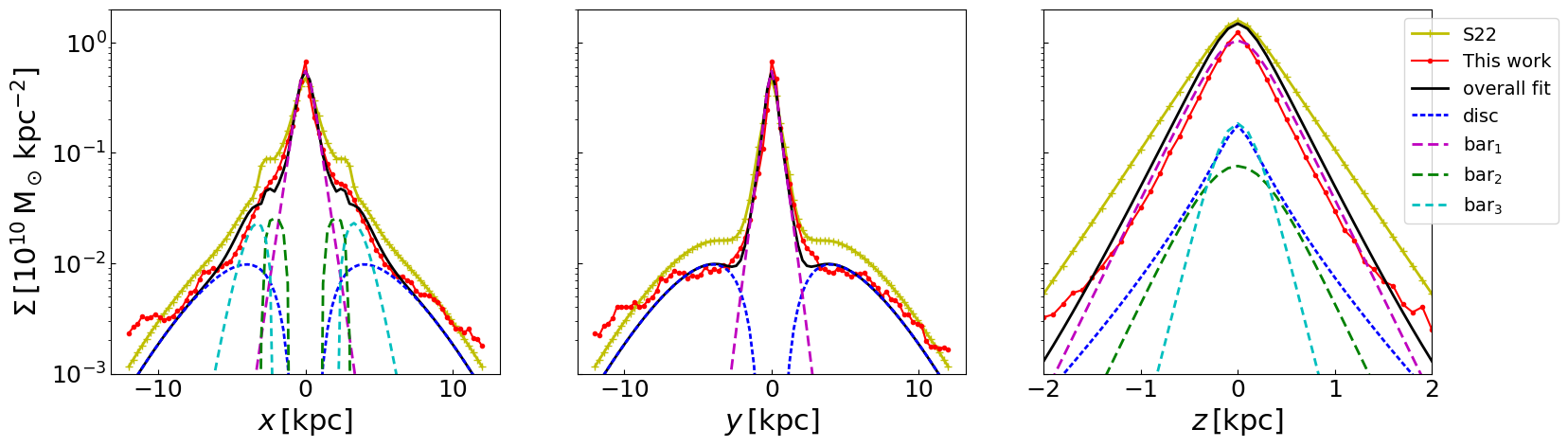}
    \vspace{-0.5cm}
    \caption{Stellar surface density profiles along the $x$ (left), $y$ (centre), and $z$ (right) axes of our model (dotted red line) compared to the analytic \protect\cite{sormani2022stellar} model (crossed yellow line) including the overall best fit of an S22-type profile to our model (shown as a black solid line) and its various components, represented by a distinct colour-coded dashed line: disc in dark blue, bar$_{1}$ in pink, bar$_{2}$ in green, and bar$_{3}$  (i.e. long bar) in light blue.}
    \label{fig:1d model}
\end{figure*}

Here we explore the stellar density distribution in the inner galaxy of our numerical model (detailed in Section~\ref{Section: Model Overview}). In particular, we conducted a detailed examination of the stellar bar/bulge structure in our model, and we compared these results against the characteristics of the stellar bar structure derived from observations of the inner Galaxy. For this comparison, we use the work of  \citet[][hereafter S22]{sormani2022stellar}, which presents an analytical description of the Milky Way's stellar bar, based on the N-body model from \citet[][]{portail2017dynamical}, developed to match observational data \citep[e.g.][]{wegg2013mapping,wegg2015structure}. The final analytical model from S22 is made out of four distinct components: an inner bulge/bar or X-shape structure (bar$_{1}$ and bar$_{2}$), a long bar (bar$_{3}$), and an axisymmetric disc. In Appendix\,\ref{appendixA}, we also include a further comparison with the stellar profiles of \citet[][hereafter R17]{ridley2017nuclear}, who use hydrodynamical simulations to model the dynamics of the central molecular zone consistent with the observed gas flows within the inner Galaxy (including the $x_2$-type orbits). That model is able to reproduce the ILR and the barred galactic component and hence also represents a good reference for the stellar profile in the Galactic centre. 

The left panel of Figure\,\ref{fig:stellar disc} shows the projected stellar surface density maps of our model in the $xy$, $xz$, and $yz$ projections, with the bar aligned with the $x$-axis. The right panel of this figure shows the residual map between our model and S22's analytical distribution. Figure~\ref{fig:1d model} then presents 1D surface-density cuts along the $x$, $y$, and $z$ axes for our model (red), the S22 analytic model (yellow), and the best-fitting S22-type decomposition to our model (black; see Appendix~\ref{appendixA}).
While our model is almost always showing lower surface densities than the S22 analytic model\footnote{Note that S22 is tailored to the central Galaxy, and as such, it should not be taken as an accurate representation of the Milky Way's disc (which is the component that dominates their $x$ and $y$ profiles beyond $\sim4-5$\,kpc). We therefore restrained our comparison to the components that dominate the central region, related to the bar.}, the differences are modest in the inner $\sim5$\,kpc: the overall shapes of the profiles match well, and the S22-type decomposition provides an adequate description of our model (black curves in Fig.\,\ref{fig:1d model}). The largest discrepancy is in the vertical direction: the $z$ cut in Fig.\,\ref{fig:1d model} (third panel) differs by up to $\sim0.3$ dex, i.e. our model can be up to a factor $\sim2$ lower in surface density. In the plane, the agreement is better: within $|x|\lesssim2$\,kpc and $|y|\lesssim2$\,kpc (left and middle panels of Fig.\,\ref{fig:1d model}), our model is typically only $\sim10$--15\% below S22. The dominant residual in the $x$ direction is associated with the bar$_2$ component around $|x|\sim3$\,kpc (green curve; Fig.\,\ref{fig:1d model}, left panel), whereas in the $y$ direction the main difference appears at $|y|\sim4$--7\,kpc where the disc component dominates in S22 (blue curve; Fig.\,\ref{fig:1d model}, middle panel), noting that this disc component is less constrained in S22.

Overall, we find that our model can be fit by a boxy/peanut bar component (bar$_{1,2}$ combined), with an approximate extension of $\sim 2$~kpc, and a long bar component (described by bar$_{3}$) with a half-length extension of $\sim3$~kpc, which aligns well with both analytical models and other observations \citep[e.g.][]{bissantz2002spiral,wegg2013mapping}. Generally, our model presents a thinner distribution than S22 and R17 models, with lower surface density within the inner $\sim 5$~kpc, and particularly so in the $z$-direction, where we find discrepancies of up to $0.3$ dex. A lower central mass could potentially impact the kinematics of the inner galaxy, which we look into in the next section.

\subsection{Stellar kinematics}
\label{sec: kinematics}

\begin{figure*}
	\centering
    \includegraphics[width=0.9\textwidth]{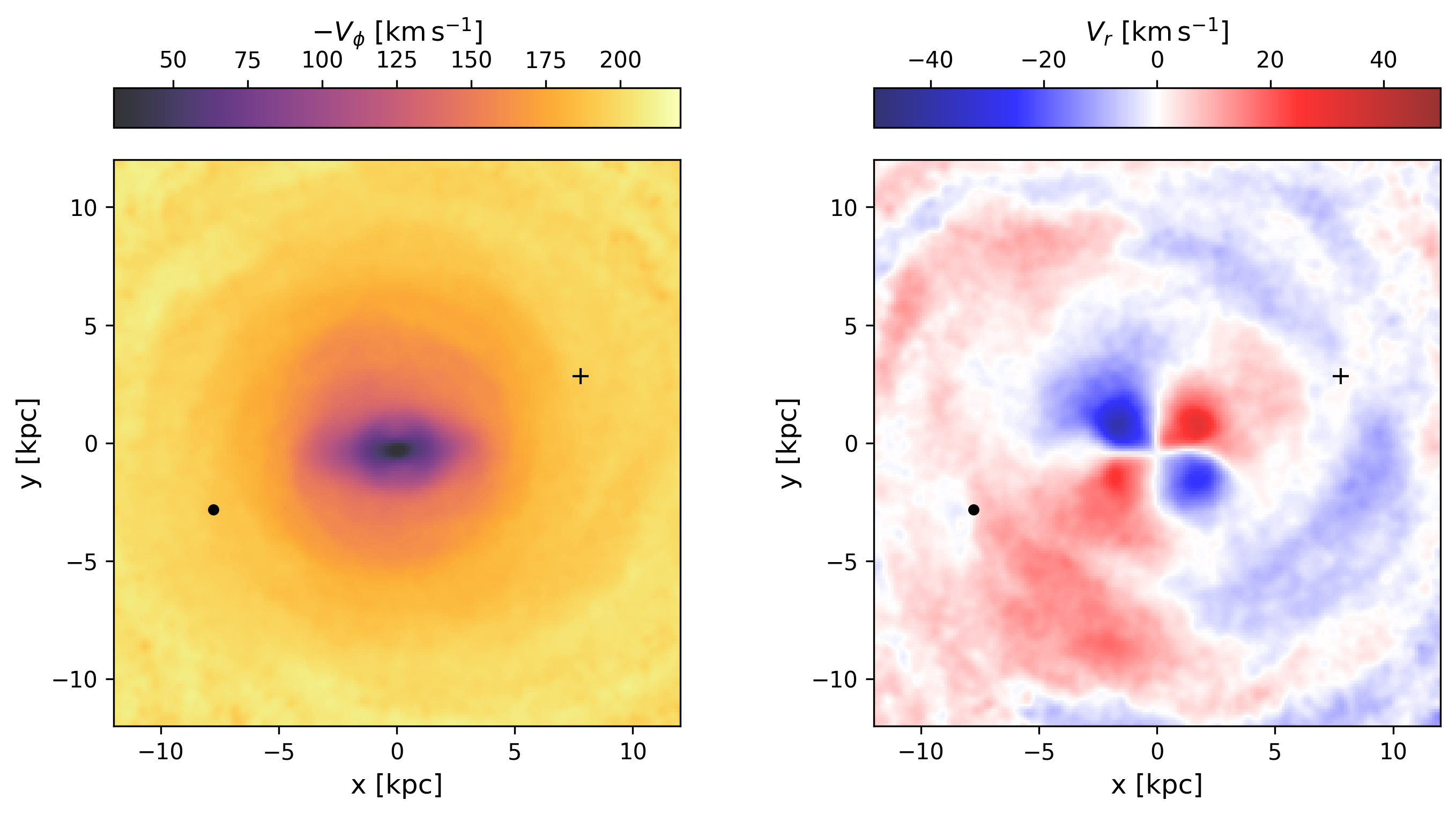}
    \caption{Top-view representations of the tangential (\(-V_\phi\), left panel) and radial (\(V_R\), right panel) velocity fields for the stellar component of our model. Note that \(V_\phi\) is negative due to the clockwise rotation of the galaxy. The bar is aligned with the $y=0$ axis, and the Sun is located at a viewing angle of $\phi_{\mathrm{obs}}=20^{\circ}$ with respect to the bar. The original Sun's position as per the best viewing angle from \citet{DuranCamacho2024} is marked as a black circle, and the Sun's reflection point (Sun$_{\rm RP}$) at $180^{\circ}$ offset is marked with a cross.}
    \label{fig:vphirzlos}
\end{figure*}

It is well known that the motions of stars and gas in galactic bars follow a quadrupole pattern, and this pattern has also been observed in the MW's Galactic centre \citep[e.g. with \textit{Gaia} DR2 and EDR3,][]{Bovy2019,queiroz2021milky}. One of the sharpest views of this kinematical pattern was revealed by \citet{drimmel2023}, who analysed the kinematics of RGB and OB stars in the Milky Way using data from \textit{Gaia} DR3. They found strong non-circular motions associated with this quadrupole pattern, with mean radial motions of the order of  $\pm 40$\,km\,s$^{-1}$. They also showed that streaming motions occur at larger galactocentric radii ($R >5$~kpc), with possible association with Lindblad resonances. For the tangential component, they observed a clear elongated feature with lower tangential velocities along the bar's major axis, within a $5$~kpc radius.

In order to benchmark our model against those observations, and understand how closely our model is able to replicate the kinematics of the centre of the MW, here we investigate the velocity fields associated with the bar pattern. In Fig.~\ref{fig:vphirzlos} we show the tangential (\(-V_\phi\), left) and radial (\(V_R\), right) velocities of our model's stellar disc, up to a radius of $12$ kpc.
In this Figure, the bar is aligned with the $y=0$ axis, and we position the Sun at a distance of $8.275$~kpc from the Galactic centre \citep[][]{GRAVITYCollaboration2021}, and such that the galactic bar has an inclination angle of $\sim 20^{\circ}$ with respect to the observer, to match the results from \citet[][]{drimmel2023}.\footnote{This angle was also found to be a suitable position for the observer in our original work \citep{DuranCamacho2024}} 

The radial velocity component in the right panel of Fig.~\ref{fig:vphirzlos} exhibits a quadrupole pattern within the long-bar extent, i.e. $\sim5$\,kpc from the Galactic centre. This quadrupole arises from bar-driven non-circular streaming motions: the sign of $V_R$ alternates across the bar major and minor axes ($y$=0 and $x$=0 lines respectively in Fig.~\ref{fig:vphirzlos}) as stars follow elongated bar-supporting orbits, producing approaching/receding (inwards/outwards) flows within the bar region. The radial velocities reach values of $\pm 45$\,km\,s$^{-1}$, which is remarkably consistent with observed findings. As shown on the left panel of Fig.~\ref{fig:vphirzlos}, the tangential velocity component also reveals an extended feature along the bar's major axis with lower velocities, as expected, given that the stars are in $x_1$ orbits, with circular velocities closer to the bar pattern speed. For our model, those velocities are of the order of $\sim 40-70$\,km\,s$^{-1}$ within the inner 5\,kpc, which is similar to the observations.

As mentioned in \cite{DuranCamacho2024}, our simulation does not develop the $x_{2}$ orbits typically observed in the vicinity of the central molecular zone, which has an extension of $\sim 250$\,pc around the Galactic centre \citep[see e.g.][]{Henshaw2016}. These orbits are oriented perpendicular to the Galactic bar and are expected within the ILR for non-axisymmetric potentials \citep[e.g.][]{Contopoulus1989,athan92a}. They emerge from the interplay between the bar's dynamics and the stellar and mass profile of the bulge. These type of orbits may not develop if there is a lack of ILRs, a lack of mass in the inner regions \citep[e.g.][]{Hasan1993,Athanassoula2013}, or the presence of a very strong bar potential \citep[e.g.][]{Contopoulus1989,athan92a}. The existence of $x_{2}$ orbits is also sensitive to the pattern speed of the bar, which determines the position of the ILRs. Faster bars push the ILR inwards, while slower bars place the ILR farther out, allowing for a more extended $x_{2}$ region \citep[see e.g.][]{athan92a,sellwood1993}. Given that our model's pattern speed  ($30 \pm 0.2$ km s$^{-1}$ kpc$^{-1}$) is comparable to the observed values \cite[$30-40$ km s$^{-1}$ kpc$^{-1}$, see e.g.][]{drimmel2023,clarke2022}, and that within our suite of models from \citet{DuranCamacho2024}, some of the simulations that do develop $x_{2}$ orbits have very similar pattern speeds, we conclude that the pattern speed is not the defining factor in this case. In addition, in \cite{DuranCamacho2024}, we demonstrated that our model does feature ILRs at $\sim$0.2 and $\sim$1.1\,kpc, and thus that is not the reason the orbits do not develop.

\begin{figure}
	\includegraphics[width=\columnwidth]{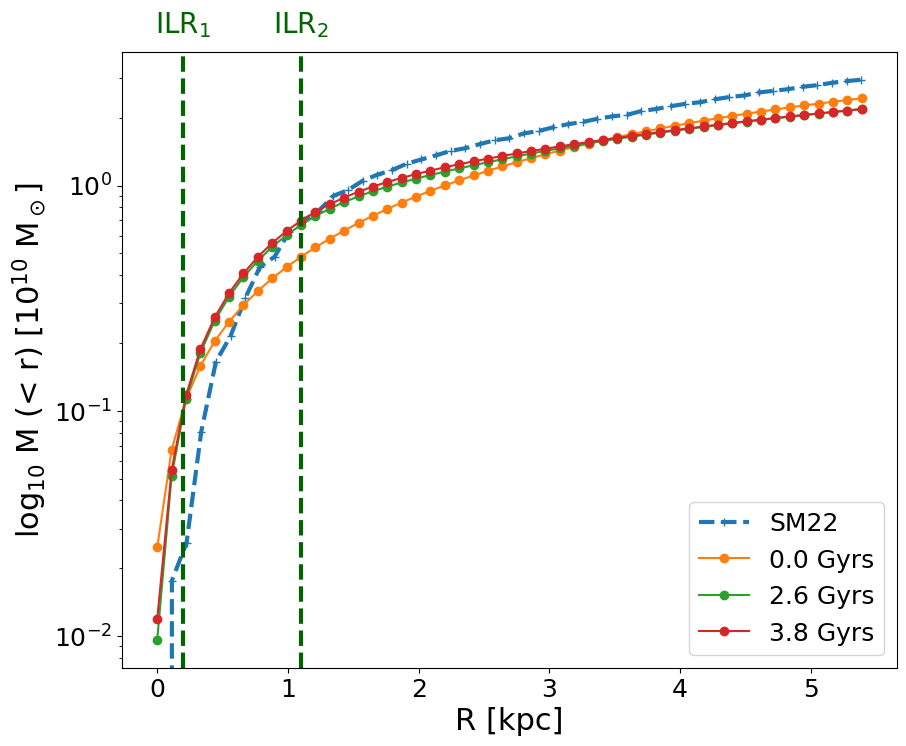}
    \caption{Spherical enclosed mass as a function of galactocentric radius for the analytical model from \protect\cite{sormani2022stellar} (blue-dashed line) and our model at initial (dotted orange line), optimal (dotted green line), and late (dotted red line) times. The position of the two ILRs in our model are shown as dashed vertical green lines.}
    \label{fig:enclosed mass spherical}
\end{figure}

A sufficiently high central mass concentration is known to contribute to the support of $x_{2}$ orbits \citep[][]{athan92a,binney1991understanding,Hasan1993}. In order to understand if the lack of $x_{2}$ orbits in our model could indeed be a result of mass missing in the inner parts of the model (as potentially hinted by Sect.\,\ref{subsection: inner Galaxy - morphology}), here we investigate the total (spherically) enclosed stellar mass as a function of galactocentric radius $R$, for our model at different times, as well as the S22\footnote{ Note that although here we show just the comparison to S22, the results would be near identical if using the R17 model instead, given that the two profiles are a close match to each other, for $R_{\rm gal} < 2.5$\,kpc (in $xy$), and $|z| < 1$\,kpc (see Fig.\ref{fig:1d R17}).} analytical model (Fig.\ref{fig:enclosed mass spherical})

This figure shows that, between the initial snapshot and the optimal time, there was some migration of mass towards the centre of the galaxy in our model, but that it stabilised at those values when compared to the later time. 
At these later times, when compared with the analytic profile, our model has, in fact, slightly greater mass in the central parts, with a discrepancy of only $\sim 3 \%$ at the second ILR. As we move outwards, this trend reverses, where our model then starts to have lower enclosed masses compared to the analytic profile, but the difference is never larger than $7 \%$ up to $5$~kpc.

In the previous section we found the largest discrepancies in the surface density distributions in the $z$ direction, and hence we also investigate the enclosed mass as a function of vertical direction $z$ (Appendix\,\ref{appendixA}). We find that in the mid-plane (within $|z|$ < 250\,pc), our model is not missing any mass until well beyond the second IRL, and that indeed it is only at higher $z$ distances that the differences become noticeable. Given that $x_{2}$-type of orbits develop in the mid-plane \citep[e.g.][]{athan92a,pichardo2002}, it is unlikely that the lack of stellar mass in the inner galaxy of our models is the main contributing factor to the absence of these orbits. This is also corroborated by the fact that despite the similarities in the S22 and R17 stellar profiles for $R < 2.5$\,kpc, the hydrodynamical simulations from \citet{Portail2015} from which the S22 profile is based, also do not develop the $x_2$ orbits, while those from R17 do \citep[see also][]{sormani2019geometry,tress2020-CMZ}.

An alternative explanation could thus be linked to the strength of the galactic bar itself. \cite{athan92a} demonstrated that the prevalence of these orbits diminishes as the strength of the bar increases. Consequently, it could be that our model's bar, by being effectively “narrower” than the observed one, might mean that the relative bar strength in our simulation is sufficiently high to suppress the formation of $x_{2}$-type orbits. It is possible that the inclusion of stellar and supernova feedback into these models will help steer up the dynamics of the gas and stars and help diminish the relative strength of the bar potential - thus potentially facilitating the development of these $x_{2}$-type orbits. This will be investigated in follow-up work. In conclusion, despite the lack of the inner $x_{2}$-type orbits in our model (affecting the central $\sim$1\,kpc), the overall density structure and kinematics of the inner regions of our simulation are close to the observed properties, making our model a suitable representation of the inner MW.

\section{Galactic disc}
\label{sec:Galactic_disk}

Understanding the Milky Way’s spiral structure is central to interpreting its dynamical state and formation history, and yet there are significant uncertainties, particularly regarding the number, shape, and longevity of its spiral arms. In this section, we investigate the spiral pattern in the disc of our simulated galaxy by examining both the stellar and gaseous components. Following this, we focus on the solar neighbourhood, exploring how our model compares to {\it Gaia} observations, and how global spiral patterns manifest locally in stellar and gas kinematics.

\subsection{Spiral pattern}
\label{sec: spiral pattern}

As mentioned in Section~\ref{sec:intro}, the configuration of the Milky Way's spiral structure remains a subject of active debate. Whether our Galaxy is composed of a two-armed, four-armed, or more complex multi-armed pattern is still unclear. In this subsection, we investigate the spiral pattern in our model’s stars and gas, aiming to shed some light on the origin of observed inconsistencies in the Milky Way.

We used Fourier transforms for this purpose, as previously applied in nearby galaxies observations \citep[e.g.][]{Elmegreen1989,Rix1995,Fuchs1999} and numerical simulations \citep[e.g.][]{Bottema2003}. By decomposing the observed structures of the galaxy into their harmonic components, this method serves as an indicator of the multiplicity/periodicity of structures in a galactic disc. The interpretation of the results from this method often relies on the assumption that the galaxy's spiral structure can be approximated by a periodic function in the azimuthal direction, with the Fourier coefficients providing a direct link to the underlying symmetry and number of arms. More details on this technique can be found in Appendix~\ref{appendix: fourier}.

\begin{figure}
	\includegraphics[width=\columnwidth]{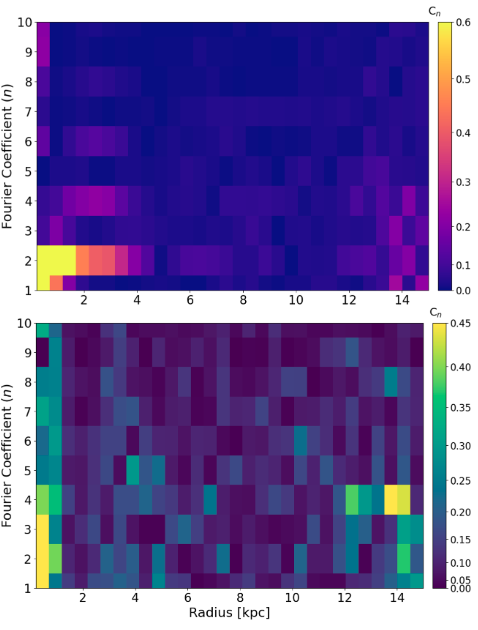}
    \caption{Density plots of the Fourier amplitudes ($C_n$) for each mode $n$ as a function of galactocentric radius for stellar particles (top) and gas cells (bottom). The colour scale represents the amplitude and therefore the strength of different spiral modes, with strong peaks indicating dominant spiral arms.}
    \label{fig:number spirals}
\end{figure}

The final representation of our model's Fourier amplitudes, $C_n$, of each Fourier coefficient (with harmonic number $n$), is shown in Fig.~\ref{fig:number spirals} as density plots across radial bins. The amplitude of each coefficient indicates the prominence of that type of multiplicity, and therefore, a prominent peak mode $n$ could be seen as directly linked to the dominant number $n$ of spiral arms at a given radius. However, as pointed out by \citet[][]{Elmegreen1989}, caution should be taken when interpreting the dominant mode as indicative of the exact number of spiral arms, as there is some degeneracy. As such, we use this Fourier decomposition as a means to inter-compare the multiplicity of the stellar and gaseous distributions, rather than to quantify their exact number of arms.

\begin{figure}
	\includegraphics[width=\columnwidth]{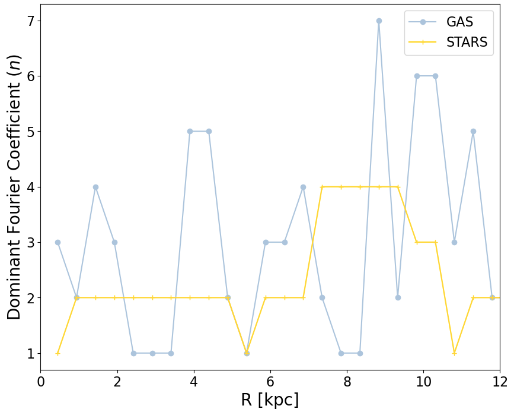}
    \caption{Dominant number of spiral arms as a function of galactic radius, obtained from the maximum Fourier coefficient (see Appendix~\ref{appendix: fourier}).}
    \label{fig:Fourier peaks}
\end{figure}

For the stellar component (Fig.~\ref{fig:number spirals}, top panel), we find a dominant $n=2$ configuration extending up to $\sim 4$\,kpc, with the highest amplitudes near the bar. Beyond that point, the pattern weakens and transitions to multiple components as the stellar potential becomes more diffuse, with no particular dominant mode. This is also illustrated in Figure~\ref{fig:Fourier peaks}, where we show the dominant mode $n$ as a function of radius, and we can see that the stars maintain a dominant $n=2$ mode up to $\sim4$\,kpc before becoming more chaotic. These findings align with observations in external galaxies, where inner discs often show 2-arm structures, while outer regions display a more flocculent nature. We also detect a much weaker $n=4$ mode, which could be indicative of a 2+2 arm pattern, where two strong arms are accompanied by two weaker arms, consistent with \textit{Gaia} studies suggesting a similar Milky Way structure \citep{Poggio2021}.

The gaseous component (Fig.~\ref{fig:number spirals}, bottom panel) reveals a more complex picture. The gas exhibits no dominant mode over extended radii, highlighting its chaotic and transient nature under a live potential, indicative of a dynamic process of formation and destruction over time. This is also evident in Figure~\ref{fig:Fourier peaks}, where the dominant mode for the gas fluctuates almost randomly between $n=1-5$ inside $\sim8$\,kpc, with complexity increasing outwards.

These findings are consistent with the spiral patterns inferred from the spiral-ridge extraction applied to the sharpened top-down stellar and gas surface density maps (Fig.\,\ref{fig:Unsharp}). In the stellar component, the ridges are low-contrast and comparatively smooth (and therefore not particularly prominent by eye in Fig.~\ref{fig:stellar disc}); nevertheless, in any radial cross section we identify $\sim2$--4 stellar arm segments over $R\simeq3$--12\,kpc. In the gas, however, the spiral pattern is sharper, in line with observations of external galaxies, where spiral structure is often more apparent in the gas/dust, and young stars than in the old stellar disc \citep[e.g.][]{Elmegreen1987, Elmegreen2011}. That said, the gaseous spiral pattern is also more fragmented: we typically identify $\sim5$--8 gaseous ridges per radial cross section over the same $R\simeq3$--12\,kpc range, indicating a segmented, non-coherent spiral morphology rather than a single large-scale grand-design pattern (see also Fig.\,\ref{fig: zoom_region}).

Overall, our model does not conform to a grand design spiral archetype. Instead, while a spiral pattern is always present, it is mostly transient, with the gas and stars often out of sync. This mirrors observations of external galaxies, where gaseous spirals display greater variability due to the ISM’s dynamical response \citep[see e.g.][]{Elmegreen1987,Maschmann2024,Dobbs2008,Renaud2013}. Such hybrid behaviour highlights the diversity of spiral structures under evolving galactic conditions \citep{Foyle2010, Hart2017}.

As noted in Sect.~\ref{sec:intro-disc}, studies suggest the Milky Way shows both grand design and multi-armed features. For example, \cite{drimmel2001} and \cite{churchwell2009} found a two-arm structure in the inner Galaxy, while the outer disc is more complex. Spitzer GLIMPSE observations confirm this duality, showing dominant and fainter arms \citep{benjamin2005}. Gas surveys \citep{Nakanishi2003, kalberla2009hi} also reveal significant variation, consistent with our model. This suggests that the Milky Way’s spiral structure may be inherently transient, with gas responding in a complex, non-density-wave-like manner, a point we explore further in the following sections (\ref{Subsec: stellar disc kinematics} and \ref{Subsec: gas disc kinematics}).

\subsection{Solar neighbourhood}
\label{Subsec: stellar disc kinematics}
\label{subsec: outer disc}

The recent work by \cite{khanna2023measuring} has identified significant streaming motions for stars within $3.5$~kpc from the Sun, demonstrating tentative correlations between the velocity structures and the spiral arms. Using an axisymmetric model to subtract from the observed velocities, they quantified overall velocity trends and found that gradients in the tangential and radial components (\(\Delta (-V_\phi\)) and \(\Delta V_R\)) were not uniform, with variations averaging $\sim 5$\,km\,s$^{-1}$ and $\sim 10$\,km\,s$^{-1}$ respectively.
However, the extent to which such velocity patterns can be related to the dynamics of the stellar and gaseous spiral pattern in the MW is incredibly hard to ascertain from observations. Indeed, as mentioned in Section~\ref{sec:intro-disc}, our knowledge of the position and extent of the stellar and gaseous arms remains highly uncertain, making these results tentative at best.

In order to shed some light into how the observed stellar kinematics pattern from \textit{Gaia} may inform us of the underlying spiral arm structure and kinematics in the solar neighbourhood, here we investigate the equivalent stellar velocity patterns in our model, and then we compare that to the underlying stellar and gaseous spiral pattern. For that, we obtain the velocity residuals in a similar way to the observations by \citet{khanna2023measuring} by contrasting the actual velocity fields observed in our simulation against an idealised rotation curve (see Appendix\,\ref{appendix: heliocentric kinematics} for more details). 

\begin{figure}
\includegraphics[width=0.48\textwidth]{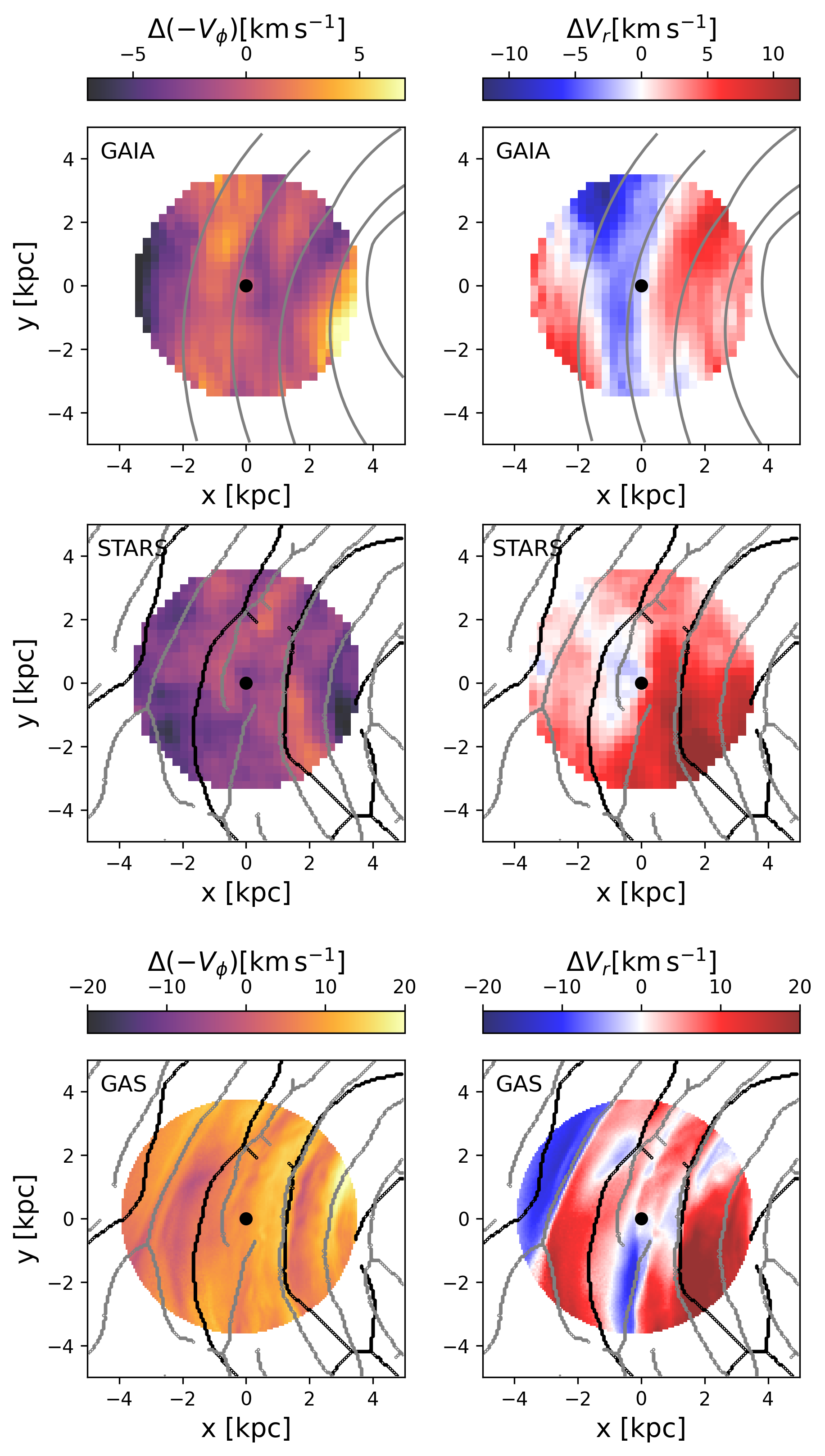}
    \caption{Heliocentric velocity residual maps for \(\Delta (-V_\phi)\) (left) and \(\Delta V_R\) (right) for \textit{Gaia} DR3 data from \protect\cite{khanna2023measuring} (top) and our model's stellar (middle) and gaseous (bottom) components in the galactic mid-plane (|\(z\)|< 0.25 kpc). The stellar maps use a binning identical to those of the {\it Gaia} data, of $250\times250$\,pc$^{2}$ pixels, for a direct comparison, while the gas maps have been binned to cells of $60\times60$\,pc$^{2}$. The solid grey lines on the top panels represent the gaseous spiral arm tracks from \protect\cite{reid19}. For our simulation, the stellar and gaseous spiral arm ridges extracted from the sharpened images of Fig.~\ref{fig:Unsharp} are superimposed in black and grey, respectively. The position of the Sun is marked as a black circle, and the Galactic centre is outside this image, at (8.275,0)\,kpc.}
    \label{fig:1d velocity res - gaia stars and gas}
\end{figure}

In Figure~\ref{fig:1d velocity res - gaia stars and gas}, we present the heliocentric velocity residuals, \(\Delta(-V_\phi)\) and \(\Delta V_R\), for the data from the \textit{Gaia} DR3 study by \cite{khanna2023measuring} (top), alongside our model's stellar (middle) and gaseous (bottom) components within the mid-plane (|\(z\)| < 0.25 kpc). Note that the stellar and gaseous panels use different colour-bar ranges to reflect their different dynamic ranges. On the \textit{Gaia} data maps we include the spiral arm tracks from \cite{reid19}, and in the simulation panels, we have included the stellar and gaseous spiral arm ridges (in black and grey respectively), as identified using an unsharpening technique (see full details in Appendix~\ref{appendix: spiral ridges}). The gaseous ridges in the simulation broadly follow the \citet{reid19} tracks, and the general trends for the stellar kinematics in our model seem to be similar to those observed by \cite{khanna2023measuring}, are qualitatively similar to the Gaia maps. The Sun lies near a transition between inward and outward motions in $\Delta V_R$ and in a region of moderately slower tangential velocities. The residual amplitudes are slightly larger in the model, reaching $\sim15$\,km\,s$^{-1}$ in $\Delta(-V_\phi)$ and $\sim25$\,km\,s$^{-1}$ in $\Delta V_R$.  Overall, the stellar kinematics in our simulation show a relatively weak correlation with the positions of the spiral ridges: neither $\Delta(-V_\phi)$ nor $\Delta V_R$ show a systematic alignment with the stellar (black) or gaseous (grey) ridges across the map.

When we looked at the velocity pattern of the gas in our model, however, things appeared a bit different. Firstly, the amplitudes of the velocity residuals are twice as large as those of the stars, reaching $\sim 30$\,km\,s$^{-1}$ and $50$\,km\,s$^{-1}$ for the tangential and radial components, respectively. The velocity changes are also a lot sharper around the position of the gaseous spiral arms. Indeed we observed that the circular velocities of the gas tend to be lower within (or close to) the gaseous spiral arm ridges, which suggests a deviation from the circular rotation curve induced by the acceleration towards the bottom of the potential well. When looking at the radial velocity component, we tend to see a number of gaseous spiral arm ridges at the convergence between inward and outward motions of the gas, which can help build up mass along the spiral arm ridges and is thus conducive to spiral arm growth. There are, however, sections of spiral arms whose velocity pattern is no longer purely converging, and it is possible that those sections of spiral arms are in the process of dissolving.

In order to investigate how fast these kinematical patterns might change, we replicated our analysis by placing the observer at different locations and times (see Appendix\,\ref{appendix: heliocentric kinematics}). The results from that exercise indicate a rapidly evolving system, with the velocity patterns changing in $\leq10$\,Myr, highlighting the transient nature of the observed streaming motions and spiral structures at any one time

Overall, our analysis shows that the pattern in the stellar structure and kinematics of our model at the chosen time is generally mild and consistent with \textit{Gaia} observations of RGB stars, but that it translates into a gaseous pattern that is much sharper. This is also consistent with observations from \citet[][]{drimmel2023} who show that OB stars (which by being younger, are potentially better tracers of the underlying gas distribution that originated them) do align better with the gaseous spiral structures, specially towards the inner disc, but with low number densities their maps remain uncertain. Given the consistency between our results and MW observations, our model could therefore offer some insight into how the formation (and evolution) of the gaseous spiral structures of the MW, albeit potentially rapidly evolving, might be intricately inter-connected with the observed kinematics of the gas, and thus we explore this correlation further in the next section.

\section{Correlation between kinematics and large-scale structures}
\label{Sec:kinematics_large_structures}

Motivated by the fact that some of the sharp changes in the gas velocity patterns found in our model seem to  typically be associated with the positions of gaseous spiral ridges, in this section we track the velocity changes both across and along spiral arm segments in order to identify what signatures are systematic, investigate how they evolve with time, and explore the mechanisms driving gas accumulation and dispersal in the galactic disc.

\subsection{Radial cross-sections of the galactic disc}
\label{Subsec: gas disc kinematics}

In this section we investigate how the radial and tangential velocity variations relate to the gas surface densities, by looking at a radial\footnote{For this exercise, we opted for a radial slice across the galaxy, such that we can see the changes of velocities across spiral arms, i.e. perpendicular to the arms' axes (which are mostly tangential in direction). In the following section, however, we also look at the changes along a given spiral arm.} cross-section of the galactic plane of our model. We define this cross-section by including all gaseous and stellar particles within the mid-plane of the galaxy (i.e. within $|z| < 0.25 $\,kpc) and located within 50\,pc of the line connecting the Galactic centre and the Sun. Our analysis focuses on particles positioned to the left of the Galactic centre (see white arrow on Fig.~\ref{fig: as drimmel} of Appendix~\ref{appendix: heliocentric kinematics}).

\begin{figure}
        \includegraphics[width=\columnwidth]{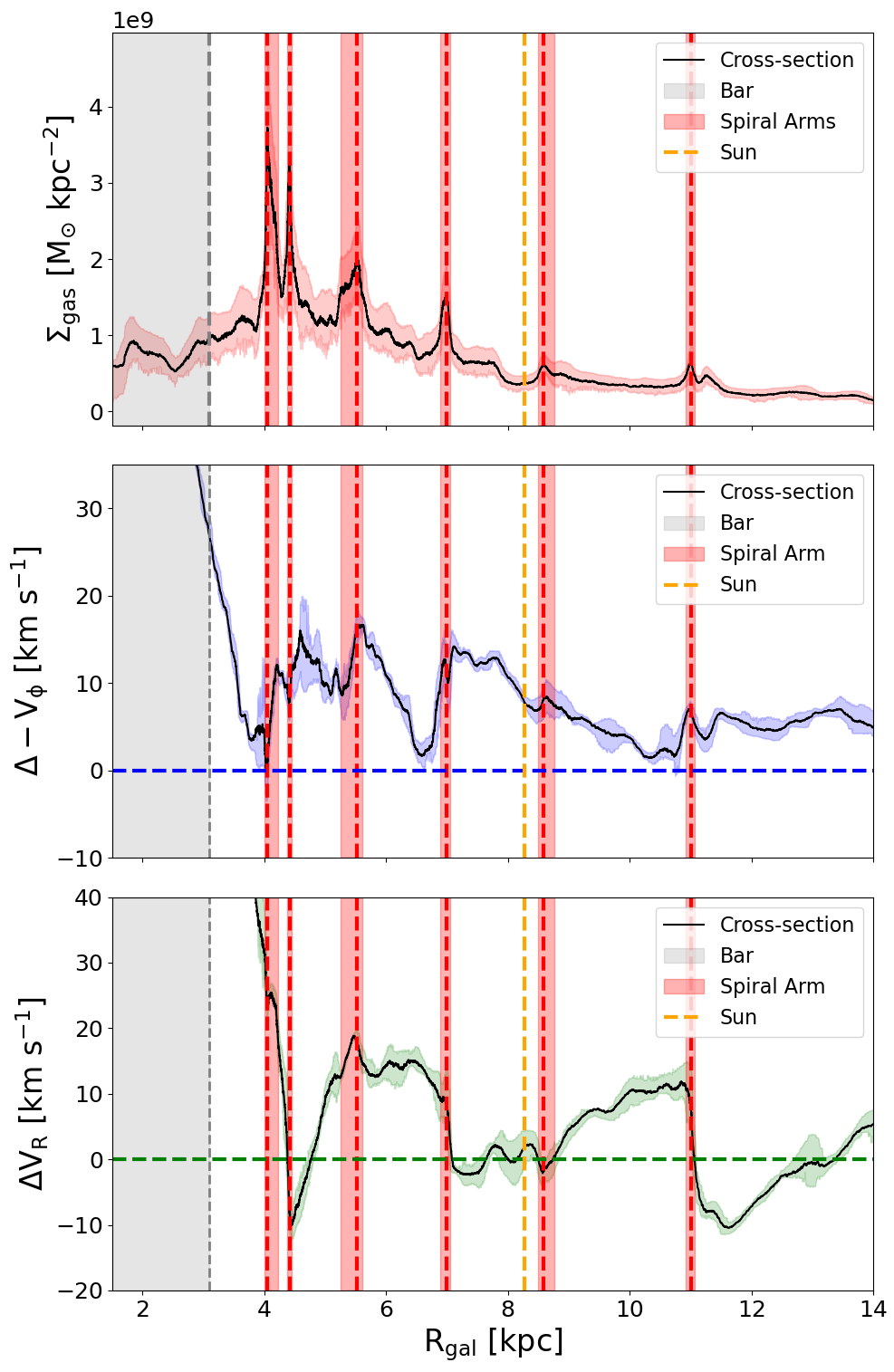}
    \caption{Surface density and velocity residuals of our model as a function of galactocentric radius for the stellar and gaseous components across the Sun's line of sight with origin at the Galactic centre. Panels show (top to bottom) the stellar surface density, gas surface density, and tangential and radial gas velocity residuals. Running medians and interquartile ranges are shown for each of the distributions, with vertical shaded regions indicating the positions of the bar (grey) and spiral arms (red), and the solar position marked by a dashed vertical orange line.}
    \label{fig:LOS_0deg}
\end{figure}

Figure~\ref{fig:LOS_0deg} shows the profile of the gaseous surface density, followed by the tangential and radial velocity residuals along that cross-section. For all plots, we include the Sun's position as a vertical dashed orange line, as well as the bar region as a shaded grey area. As our focus is on the dynamics in and around the over-densities of spiral arms,
we excluded this central (bar-dominated) region from our analysis.

The first panel of Figure~\ref{fig:LOS_0deg} shows that the gaseous spiral arms exhibit pronounced overdensities, with up to six distinct arms identified as local peaks in the distribution at $R\simeq 4.0, 4.2, 5.8, 7.0, 8.5,$ and $11.1$\,kpc, marked by red dashed vertical lines, with their approximate widths\footnote{The peaks were identified by finding local maxima in the running median of the gas surface density, and the widths were estimated using the python \textit{peak\_width} function, which measures the full width at half maximum (FWHM) of each peak.} indicated by vertical red shaded areas. These peak radii are in good agreement with the positions where the extracted gaseous spiral ridges intersect this line of sight (Appendix~\ref{appendix: spiral ridges}) and are consistent with the ridges used in Sect.~\ref{Subsec: stellar disc kinematics}. For reference, the same arm intervals are overplotted in the lower panels, which show the tangential (middle) and radial (bottom) velocity residuals along the same slice.

In the tangential residual profile (Fig.~\ref{fig:LOS_0deg}, middle panel), the positions of the spiral arms are mostly associated with a positive slope, i.e. $\Delta(-V_\phi)$ increasing across the arm from its inner to outer edge. This corresponds to gas rotating slower on the inner side of the arm and more rapidly on the outer side, reducing the local differential rotation (shear) within the arm. Deviations from this trend occur mainly for the weakest arm peaks (e.g. at $R\simeq 8.5$). The amplitudes are largest in the inner disc ($|\Delta(-V_\phi)|\sim10$--15\,km\,s$^{-1}$ at $R\lesssim$ 8 kpc), where the spiral features are strongest, and decline at larger radii.

The third panel of Figure~\ref{fig:LOS_0deg} shows the radial velocity component as a function of galactocentric radius. In this frame, a downward slope indicates regions where gas is radially converging, regardless of whether there is a change in sign from outward to inward motion, though the most significant convergence typically corresponds to a sign change. Conversely, an upward (positive) slope suggests divergence in the gas, and a flat slope indicates stable regions, where the gas is neither radially converging nor diverging, regardless of the direction or magnitude of the overall flow. A prominent example is the sharp drop in $\Delta V_R$ at the end of the bar region (grey shaded area), with an amplitude of $\sim40$\,km\,s$^{-1}$. This is substantially larger than the bar-related change in $\Delta(-V_\phi)$ along the same slice ($\sim15$\,km\,s$^{-1}$), and it is at the end of this region of strongly converging gas that we find the strongest spiral arms in the gas surface density plot. At larger radii, most gas exhibits relatively stable outward motions, with the gaseous spiral arms located where $\Delta V_R$ shows locally negative slopes (often where $\Delta V_R$ transitions from positive to negative), with typical gradients across the arms of $\sim12$--15\,km\,s$^{-1}$\,kpc$^{-1}$. In contrast, inter-arm regions more often show positive slopes (divergence), generally with weaker gradients than those associated with the arm peaks.

In order to test if these correlations are consistent across the galaxy, in Appendix\,\ref{appendix: cross-sections} we also examine two more cross-sections positioned $\pm 20^{\circ}$ above (left) and below (right) the Sun's line of sight (see Figure~\ref{fig:LOS_20deg}). We find that all trends are similar, with spiral arms typically associated with sharp negative slopes of $\Delta V_{R}$, and sharp positive slopes in $\Delta (-V_{\phi})$, while inter-arms show positive albeit very milder slopes in $\Delta V_{R}$.

Our analysis suggests that the position and strength of the spiral arms strongly correlate with distinct velocity gradients in radial and tangential velocities. We also find that the strongest gaseous spiral arms, characterised by strongly converging gas, are mostly found in the inner galaxy, where the velocity changes are sharper. These velocity gradients occur predominantly in the inner disc, within/around the bar corotation radius ($R_{\rm CR}\simeq 6.1$ kpc for this model; see the resonance analysis in \citet{DuranCamacho2024}), where the bar-driven non-axisymmetric streaming field is expected to be strongest. These results align with the findings of \citet{Urquhart2014}, who highlight the prominence of the Sagittarius arm as a major gaseous feature in the Milky Way, based on observations of young massive stars from the Red MSX Source (RMS) survey. It remains to be seen, however, how persistent these velocity patterns are throughout time, which we explore in the following section.

\subsection{Time evolution of spiral arm segments}
\label{Subsec:time_evolution_spiral_arm}

\begin{figure}
    \includegraphics[width=0.48\textwidth]{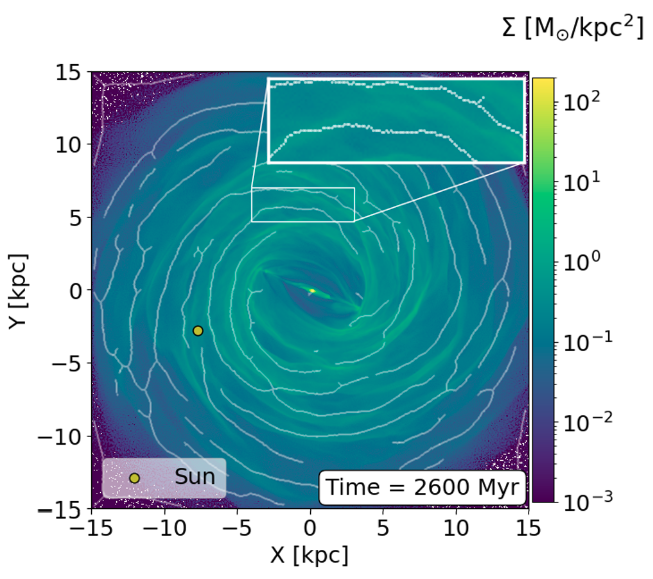}
    \caption{Top-down view of the stellar surface density distribution in the simulation at $t = 2.6$~Gyr. The inset box shows a zoom-in of the spiral arm segment examined in detail in Section~\ref{Subsec:time_evolution_spiral_arm}.}
    \label{fig: zoom_region}
\end{figure}

In this section, we use our model to explore the process of spiral arm growth and dissolution over time and its link to the underlying kinematics. We used the gaseous spiral ridge points (see Appendix~\ref{appendix: spiral ridges}) to trace the underlying skeleton of the gaseous spiral structure and follow the evolution of a selected spiral arm segment (see inset box in Fig.\,\ref{fig: zoom_region}) over a period of $\sim20$\,Myrs. This particular region was chosen as an example of an arm with a section that is actively growing, and breaks within this time frame (the time evolution of this spiral arm segment\footnote{A video of the time evolution of the velocity fields for the entire galactic disc can be found in the FFOGG project website: https://ffogg.github.io} can be seen in Fig.~\ref{fig:arm_times}).

We investigated the evolution of physical and kinematic conditions of the arm by tracking key quantities as a function of distance along the spiral ridge, $s$, namely the gas volume density ($\rho$), radial and tangential velocity residuals ($\Delta V_{\rm R}$ and $\Delta(-V_{\phi})$), and velocity divergence ($\nabla\cdot\mathbf{V}$). In addition, we also track the fraction of dense gas ($f_{\rm dense}$), defined as the percentage of mass above a volume density threshold of $700~\mathrm{cm}^{-3}$ \citep[in line with sink/star-particle insertion thresholds used in similar galaxy-scale simulations; see e.g.][]{dobbs2015,tress2020simulations}.

\begin{figure*}
        \includegraphics[width=\textwidth]{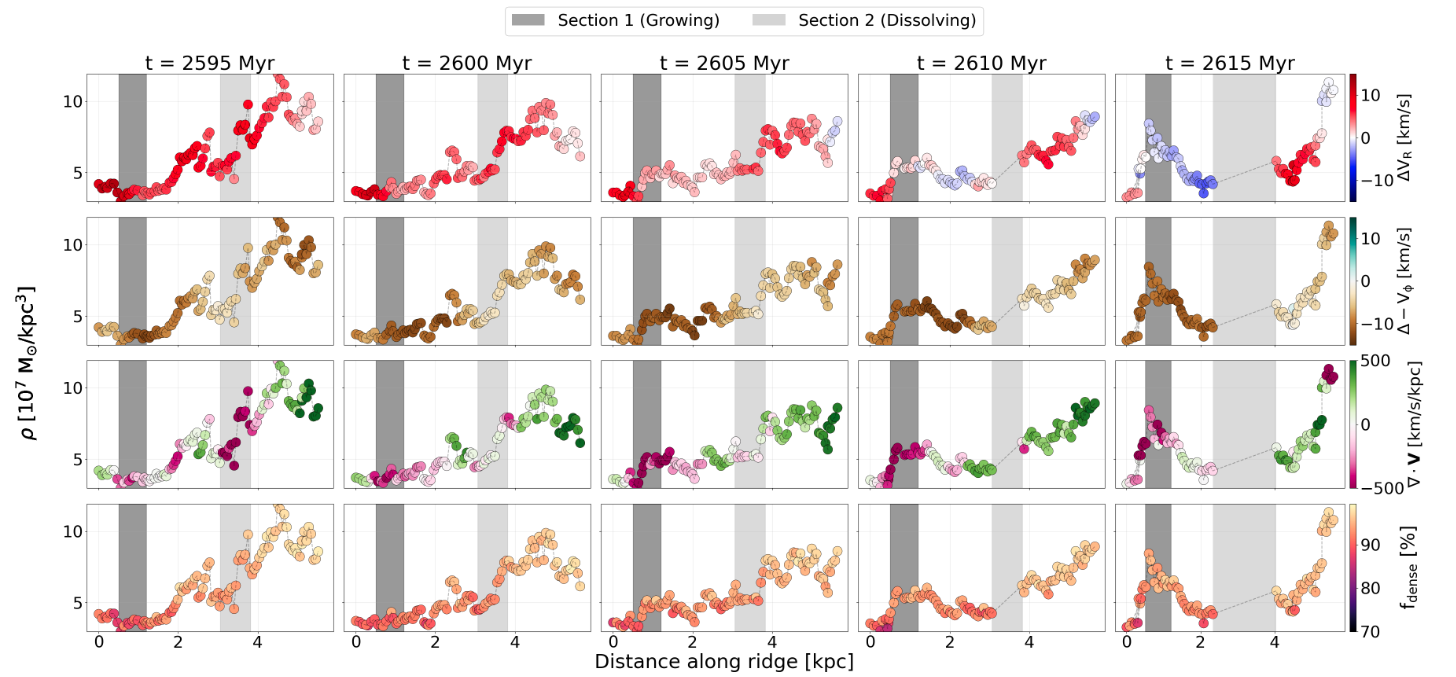}
    \caption{Time evolution of the volume density of the spiral arm segment selected in Figure~\ref{fig:arm_times} (Appendix~\ref{appendix: arm evolution}), shown as a function of distance along the spiral ridge. Each column represents a different snapshot, covering a 20~Myr period in 5~Myr intervals. The points are colour-coded (from top to bottom) by (i) radial velocity residuals (\( V_{R} \)), (ii) tangential velocity residuals (\( \Delta - V_{\phi} \)), (iii) velocity divergence (\( \nabla \cdot \mathbf{V} \)), and (iv) the fraction of gas cells exceeding a critical volume density threshold of \( 700~\mathrm{cm}^{-3} \). Negative values of \( \nabla \cdot \mathbf{V} \) indicate converging gas. The two spiral arm sections discussed in the main text are highlighted with different shades of grey and tracked across all times as a visual guide.}
    \label{fig:arm_evolution_along_ridge}
\end{figure*}

The results are presented in Figure~\ref{fig:arm_evolution_along_ridge}, where each column corresponds to a different snapshot (separated by 5\,Myr), while the four rows colour-code the ridge points by $\Delta V_{\rm R}$, $\Delta(-V_{\phi})$, $\nabla\cdot\mathbf{V}$, and $f_{\rm dense}$, respectively. Throughout this section we refer to two fixed ridge intervals (shaded in grey in Fig.~\ref{fig:arm_evolution_along_ridge}): a growing segment ({\it Section~1}) at $s\simeq$ $1.2$--$1.8$ kpc and a dispersing segment ({\it Section\,2}) starting at $s\sim$ $3.1$ kpc.

Focusing first on the growing segment, Fig.~\ref{fig:arm_evolution_along_ridge} shows that {\it Section\,1} initially appears comparatively low-density and only weakly convergent, as indicated by small negative values of $\nabla\cdot\mathbf{V}$ (third row), and that as time progresses it becomes more convergent and densities increase considerably. The kinematic evolution is consistent with this build-up, as this particular section exhibits lower $\Delta(-V_{\phi})$ values (second row) than the surrounding areas, indicating that gas is slowing down in its rotation and therefore contributing to the gathering of material.

This section also exhibits a very mild change of radial velocities (first row) along it, but given that the arm segment is nearly perpendicular to the radial direction, it suggests that this segment has a coherent ``bulk motion'', with these radial velocity gradients only mildly reshaping the structure over time. Over the 20\,Myr shown, the typical $\Delta V_{\rm R}$ values within {\it Section\,1} shift from predominantly positive (outwards) at early times to predominantly negative (inwards) at late times, illustrating that the local flow field evolves rapidly even as the ridge grows in density. This behaviour is consistent with Fig.~\ref{fig:arm_times}, where {\it Section\,1} ($x\simeq [-3.3,-0.5]$\,kpc and $y\simeq [5.4,5.8]$\,kpc) lies close to an interface between outward and inward motions, i.e. a convergence zone in the 2D field. Finally, $f_{\rm dense}$ (fourth row) increases within the same $s$-interval as $\rho$ rises, indicating that growth of the ridge is accompanied by an increasing fraction of gas above the star-forming density threshold.

In contrast, the dispersing segment shows the opposite sequence. Initially, the gas in this region has a high density peak and shows convergence, but over time it becomes progressively more divergent, eventually leading to the dissolution of the arm. This transition is also marked by a positive gradient in radial velocity residuals, as well as an increase of the tangential component, such that it is rotating faster in the leading part of this section, and hence the two ends of this section effectively feel a different `bulk motion' that contributes to pulling it apart. This process is accompanied by a redistribution of mass, as the volume density progressively decreases. Consistently, $f_{\rm dense}$ decreases as the ridge disperses. The 2D perspective shown in Fig.~\ref{fig:arm_times} supports this interpretation. At the location corresponding to {\it Section\,2} ($x\simeq [-1.6,0.1]$\,kpc), the converging region present at early times is progressively replaced by a diverging `valley' that splits the ridge into two fronts that subsequently move radially in opposite directions.

Interestingly, this alternating convergence-to-divergence pattern within spiral arm ridges over time, is visible not just in this arm, but across the entire disc. Spiral arm ridges grow on convergence zones, at the interfaces between gas flows of opposing radial velocities, but after $20-40$\,Myrs, the location of the once-dense ridges becomes divergent, as if the two shock fronts that formed it have now moved through each other. Spiral ridges thus split into two and continue travelling radially (one inwards, one outwards) until they hit another front travelling in the opposite direction. Any given spiral arm seems to ``survive'' in its densest form for $<$\,20\,Myrs before it splits, equivalent to the expected crossing time of the typical radial motions of the order of $\sim$20\,km\,s$^{-1}$, for a spiral ridge of 400\,pc width. This would suggest that the transient nature or the spiral pattern - their formation, survival, and reshaping - is intricately connected with the radial motions of the gas, a large portion of which are driven and sustained by the galactic bar and its butterfly pattern, which extends well into the galactic disc. This is obviously subject to this specific model, where we do not include self-gravity or feedback, and the lifetime of any given spiral ridge might be heavily affected by either of those factors considerably.

In summary, our combined analysis reveals a strong connection between the large-scale gas dynamics and the localised evolution of spiral arm segments in our galactic model. On global scales, peaks in the gas surface density trace the spiral arms and coincide with marked deviations in both radial and tangential velocity residuals. Gas tends to decelerate in their circular motion inside the arms, while radial velocity patterns highlight alternating converging and diverging flows at spiral arm locations. These large-scale features are consistent with previous N-body simulation results. For example, \citet{Kawata2014} found that gas around spiral arms exhibits sharp velocity variations, and \citet{baba2015dynamics} showed that spiral arms form through swing amplification and dissolve due to dynamical forces. This is also in line with recent \textit{Gaia} observations by \citet{Natsuki2024}, which show that the Perseus arm appears to be dissolving as stars escape the arm, while the Outer arm is likely growing with stars converging, both behaviours closely linked to the observed velocity fields. Our results further illustrate the inter-linked and transient nature of spiral structures and gas dynamics. While a spiral pattern is always visible and distinct in the gas, any given spiral segment only survives as a single coherent feature for a few tens of megayears, and this timescale is dictated by the amplitude of the radial motions present in the gas. Indeed, if this is the dominant mechanism by which arms grow and dissolve, then we might be able to infer the typical lifetime of spiral arms in the Milky Way, by measuring the amplitude of the gas radial motions. Of course the ability for a given gaseous spiral arm to grow and survive large-scale disruption in these dynamically evolving systems, might change if gas self-gravity and stellar feedback were to be included in our model, and we will explore this in future work.

\section{Summary and conclusion}
\label{Sec: Conclusions}

In this work, we made use of the isothermal simulation presented in \cite{DuranCamacho2024} as Model 4 to provide a good representation of the overall observed Milky Way structure and conducted an in-depth analysis of the structure and kinematical properties of both the stellar and gaseous components across various regions of that model galaxy. In particular, we investigated the influence of large-scale dynamics on defining the large-scale structures as well as their potential impact on the agglomeration and disruption of gas, which in turn could set the initial conditions for the star formation processes. Our findings can be summarised as follows:

\begin{itemize}
    \item {The stellar distribution of our model's inner galaxy (Section\,\ref{subsection: inner Galaxy - morphology}) can be well described by a boxy/peanut bar structure with a half-length of $\sim$2 kpc and a long bar component with a half-length of 3.1 kpc, aligning well with both analytical models and observational data of the Milky Way \citep{bissantz2002spiral, wegg2015structure,sormani2022stellar,ridley2017nuclear}. Nonetheless, the model exhibits a thinner surface density distribution within the inner 6 kpc, with the largest discrepancies in the vertical direction.}
    \item {The kinematic pattern of our model's inner galaxy (Section\,\ref{sec: kinematics}) shows a radial velocity quadrupole pattern and a tangential velocity deceleration region along the bar, and both have a similar extent and amplitude to the observations from {\it Gaia} \citep[e.g.][]{GaiaDR32021,drimmel2023}.}
    \item {Our model does not produce $x_2$ orbits in the inner-most regions ($<1$\,kpc), and we explored if this could be due to a lack of central mass (Section~\ref{sec: kinematics}). We find that the enclosed stellar mass within 5\,kpc is only $\sim 7 \%$ lower than that of other analytic models that reproduce the inner MW, but most of the discrepancies are for larger $z$ heights. In the mid-plane, the discrepancy within the ILRs (at $\sim$0.2 and 1.1,kpc) is negligible, suggesting that mass alone is not responsible for the absence of $x_2$ orbits. Instead, we argue that the strength of the bar and the lack of feedback likely play a more significant role in suppressing them. This hypothesis will be explored in future work.}
    \item{We examined the spiral structure of our model using Fourier decomposition and by extracting the spiral arm ridges from both the stellar and gas surface density maps (Section~\ref{sec: spiral pattern}). Both analyses suggest that the gas shows sharper and more numerous spiral arm features throughout the disc than the stars, although neither component can be described by a well-defined number of spiral arms over extended radii, and the gas and stars are not necessarily synchronised in their spiral structure. These findings highlight that although a spiral structure is always present in both components, it does not conform to a grand-design archetype. If taken as an analogue to the MW, this view of a more dynamically evolving spiral pattern in our model could explain the observational difficulty in determining the number of spiral arms in the MW.}
    \item{We analysed the stellar kinematics around the solar neighbourhood (Section~\ref{Subsec: stellar disc kinematics}) and compared them to the observed stellar velocity distribution from \textit{Gaia} DR3 \citep[e.g.][]{khanna2023measuring,drimmel2023}. We find our model to be largely in agreement with the observations (albeit with slightly larger amplitudes in the velocity residuals), but we find that there is not always a unique correlation between specific changes in the velocity of the stars and the position of the spiral arms (may those be the gas or stellar arms). However, we find that the gas experiences much sharper kinematic changes in both radial and tangential directions that correlate better with the position of the gaseous spiral arms.} 
    \item{By analysing the specific changes of velocity along cross-sections of the disc (and therefore cross-sections of various spiral arms; see Section\,\ref{Subsec: gas disc kinematics}), we found systematic patterns associated with the position of strong gaseous spiral arms. In particular, we find they are regions of reduced differential rotation (seen as increasing tangential velocities from the inner to the outer side of the arm), and regions of strong radial velocity convergence.}
    \item {Our time-evolution study of a spiral arm segment (Section~\ref{Subsec:time_evolution_spiral_arm}) suggests that these tangential and radial velocity patterns are rapidly evolving and that this leads to the growth and dissolution of spiral arm segments in relatively short timescales ($< 20$\,Myrs). We find that arm growth is preceded by converging radial flows, coinciding with increases in volume density and in the fraction of gas above a critical density threshold for potential star formation. Conversely, arm destruction correlates with diverging flows and velocity gradients that shear the structure apart. These findings support a scenario in which the persistence of spiral arms is closely related to the large-scale kinematic environment and, in particular, the amplitude of radial motions - driven in large by the extended butterfly pattern induced by the galactic bar well into the galactic disc.} 
\end{itemize}

In conclusion, our findings reveal a transient nature of the spiral pattern in our model, challenging the traditional view of these structures as stable features in galaxies akin to the MW, as predicted by the density wave theory. While our isothermal simulations do not include star formation, the observed gas dynamics suggest that spiral arms primarily act as mechanisms for gas accumulation in a more complex and chaotic fashion than the typical grand-design spiral patterns. These fluctuating patterns indicate that the relationship between spiral arms and star formation is more complex and potentially less direct than previously assumed. This relationship is likely dependent on the relative timescales involved in the formation and dissolution of spiral arms and the timescales for the gravitational collapse of molecular clouds. 

While these results suggest that the rapidly changing large-scale kinematics of the galaxy can play a significant role in shaping the ISM, self-gravity, chemistry, and stellar feedback  are all factors that will influence the star formation process itself and potentially affect the longevity and stability of spiral arms. Incorporating these physical processes in future simulations will help clarify whether gas accumulation in spiral arms under specific conditions is sufficient to initiate star formation or whether additional environmental factors play a more dominant role. This will be explored in follow-up work, along with the inclusion of ISM gas chemistry and gas self-gravity into this model. Overall, this work offers valuable insights into the processes potentially shaping the ISM in the MW, and our model thus offers a basis for a more dynamic framework to study the formation and evolution of molecular clouds and stars in our Galaxy.

\section*{Data availability}

The maps for the top-down surface density, velocity, and \textit{lv} projection of our best model have been made publicly available through the publication of \citep{DuranCamacho2024} and can be found at the website of the Following the Flow Of Gas in Galaxies (FFOGG) project (\url{https://ffogg.github.io/}). Any data not available on the website can be provided upon requests to the authors. With this article, we further make available a video of the time evolution of the spiral structure of our simulation, with specific focus on the evolution of the different kinematical patterns studied here.

\begin{acknowledgements}
    The authors thank Mattia Sormani and Shourya Khanna for insightful comments and discussions. The calculations presented here were performed using the supercomputing facilities at Cardiff University operated by Advanced Research Computing at Cardiff (ARCCA) on behalf of the Cardiff Supercomputing Facility and the HPC Wales and Supercomputing Wales (SCW) projects. EDC acknowledges funding from the European Union grant WIDERA ExGal-Twin, GA 101158446. ADC and EDC acknowledge the support from a Royal Society University Research Fellowship (URF/R1/191609).  
\end{acknowledgements}

\bibliographystyle{bibtex/aa}
\bibliography{bibtex/sorted_updated_references}

\begin{appendix}

\section{Inner Galactic stellar profiles}
\label{appendixA}

In Section~\ref{section: inner Galaxy}, we investigate the stellar surface density distribution in our model and compare it to the work from \citet[][S22]{sormani2022stellar} and \citet[][R17]{ridley2017nuclear}. Here we provide more details on that comparison, as well as the fitting of the S22-type of profile to our model.

\cite{sormani2022stellar} presents an analytical model tailored to describe the stellar profile of the inner Galaxy, based on the N-body model from \citet[][]{portail2017dynamical}, which had been developed to match observational data \citep[e.g.][]{wegg2013mapping,wegg2015structure}. Their global density profile (Eq.~\ref{eq. profile}) includes 4 components: a boxy/peanut region in the centre which is described by the components bar$_{1,2}$, a long bar described by bar$_{3}$, and an axisymmetric disc. 

\begin{equation}
\label{eq. profile}
\rho(x, y, z) = \rho_{\text{bar,1}} + \rho_{\text{bar,2}} + \rho_{\text{bar,3}} + \rho_{\text{disc}}
\end{equation}\\

The profile for the first barred component is described by
\begin{equation}
\label{eq. bar 1}
\rho_{\text{bar,1}}(x, y, z) = \rho_1 \, \text{sech} \left( a^m \right) \left[ 1 + \alpha \left( e^{-a_+^n} + e^{-a_-^n} \right) \right] e^{-\left( \frac{r}{r_{\text{cut}}} \right)^2}
\end{equation}

\noindent where

\begin{equation}
a = \left\{
\left[ \left( \frac{|x|}{x_1} \right)^{C_{L,1}} + \left( \frac{|y|}{y_1} \right)^{C_{L,1}} \right]^{\frac{C_{L,1}}{L}} + \left( \frac{|z|}{z_1} \right)^{C_{U,1}} \right\}^{\frac{1}{C_{U,1}}},
\end{equation}
\begin{equation}
a_{\pm} = \left[ \left( \frac{x \pm c z}{x_c} \right)^2 + \left( \frac{y}{y_c} \right)^2 \right]^{\frac{1}{2}},
\end{equation}
\begin{equation}
r = \left( x^2 + y^2 + z^2 \right)^{\frac{1}{2}}.
\end{equation}\\

The profiles for barred components i $ = \{2, 3\}$ are given by
\begin{equation}
\label{eq. bar 2,3}
\rho_{\text{bar,i}}(x, y, z) = \rho_i \, e^{-a_i^{n_{i}}} \text{sech}^2 \left( \frac{z}{z_i} \right) e^{-\left( \frac{R}{R_{\text{i,out}}} \right)^{n_{i,\text{out}}}} e^{-\left( \frac{R_{\text{i,in}}}{R} \right)^{n_{i,\text{in}}}},
\end{equation}

\noindent where 

\begin{equation}
a_i = \left\{
\left[ \left( \frac{|x|}{x_i} \right)^{C_{L,i}} + \left( \frac{|y|}{y_i} \right)^{C_{L,i}} \right]^{\frac{C_{L,i}}{L}} \right\}^{\frac{1}{C_{L,i}}},
\end{equation}

\begin{equation}
R = \left( x^2 + y^2 \right)^{\frac{1}{2}}.
\label{eq radius}
\end{equation}\\

Finally, the galactic disc is represented with an axisymmetric profile as defined by 
\begin{equation}
\label{eq. disc}
\rho_{\text{disc}}(R, z) = \frac{\Sigma_0}{4 z_d} e^{-\left( \frac{R}{R_d} \right)^{n_d}} e^{-\frac{R_{\text{cut}}}{R}} \text{sech} \left( \frac{|z|}{z_d} \right)^{m_d},
\end{equation}

\noindent where $R$ is the cylindrical radius from equation~\ref{eq radius}. We note however that their analytical model was produced by fitting data from the inner Galaxy, and hence this disc profile should not be taken as an accurate representation of the Milky Way's disc beyond $\sim4-5$\,kpc.

\begin{figure*}
	\includegraphics[width=\textwidth]{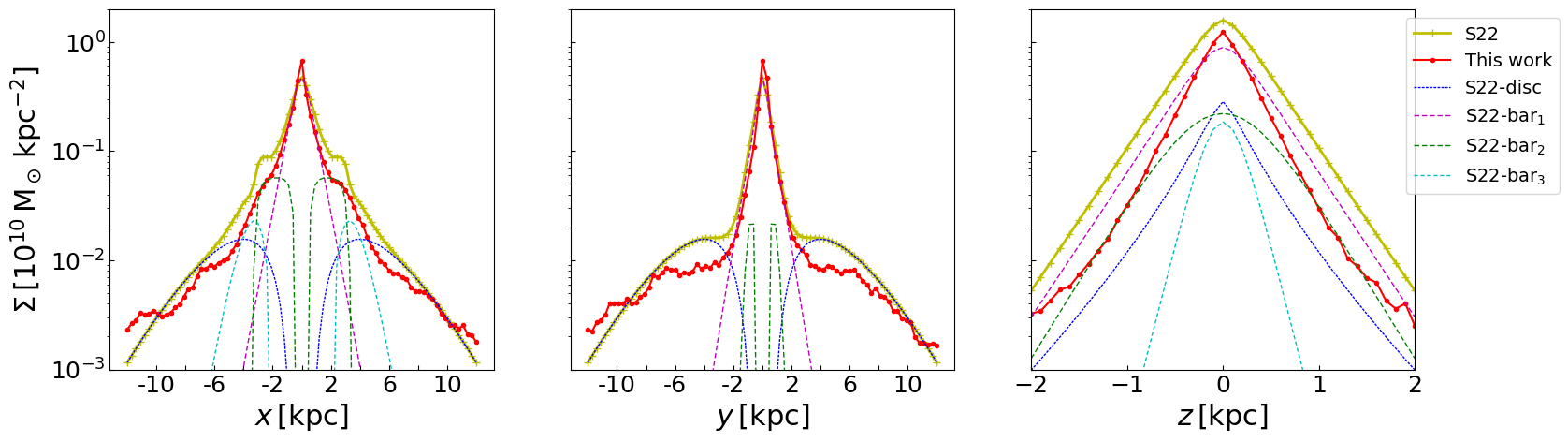}
    \includegraphics[width=\textwidth]{plots_memory/1d_surface_dens_ModelMstar425_BulgeFraction10_CC9_SM22_fitting.png}
    \caption{Stellar surface density along the $x$ (left), $y$ (middle), and $z$ (right) axes of our model compared to the \protect\cite{sormani2022stellar} analytic model (S22). Top row shows the comparison between our model (dotted red line) and the S22 model (crossed yellow line). The different dashed coloured lines represent the various galactic components of the S22 model: disc in dark blue, bar${1}$ in pink, bar${2}$ in green, and bar$_{3}$ or long bar in light blue. Bottom row includes the overall best fit of our model (black solid line) along with its components, represented by the same colour-coded dashed lines as in figures in the top row. Left-hand and middle panels correspond to the $x$ and $y$ axes in the ($x$,$y$) top-view map, while right-hand panels corresponds to the $z$-axis in the ($x$,$z$) map.}
    \label{fig:1d SM22}
\end{figure*}
\begin{figure*}
	\includegraphics[width=\textwidth]{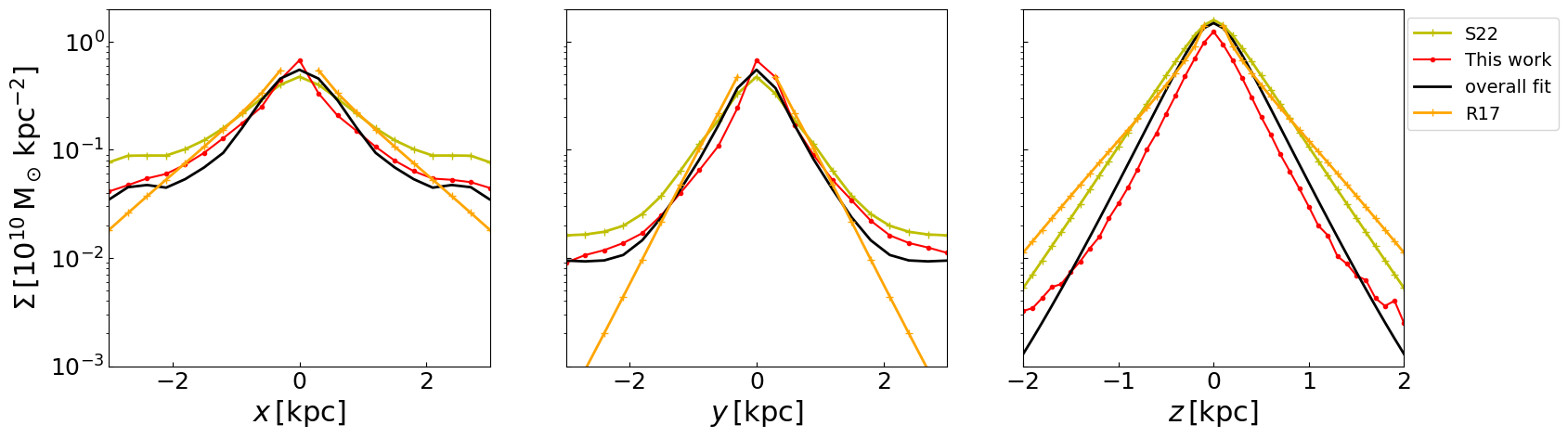}
    \caption{Similar to Fig.~\ref{fig:1d model}, but zoomed-in to the $\pm 3$ kpc region in the $x$ and $y$ directions and including the profile from R17 excluding the halo contribution (in orange), the S22 model in yellow, our model in red, and the S22-type profile fit to our model in black.}
    \label{fig:1d R17}
\end{figure*}

Table~\ref{table 2} shows the $27$ different parameters as adopted by S22, as well as the parameters which were further fitted to our model's stellar distribution, using the same profiles. Note that the most notable differences of our fit compared to S22's, are for the barred component 2 (which controls the inner bar's boxy appearance), where both the size and peak surface densities are changed. Figure\,\ref{fig:1d SM22} shows the resulting 1D profiles of S22 when slicing the surface density images through the three principal axes as well as our model's profile and respective fit.

\begin{table*}
\centering
\caption{Best-fitting parameters for our numerical model.}

\begin{tabular}{lccc|c cccc}
\hline
\textbf{Parameter} & \textbf{This work} & \textbf{S22} & \textbf{Units} & & \textbf{Parameter} & \textbf{This work} & \textbf{S22} & \textbf{Units} \\
\hline
\\
\multicolumn{4}{c}{\textbf{\underline{Barred component 1}}} & & \multicolumn{4}{c}{\textbf{\underline{Disc}}} \\
\\
$\rho_1$ & 0.386 & 0.316 & $10^{10} \mathrm{M}{\odot} \mathrm{kpc}^{-3}$ & & $\Sigma_0$ & 0.065 & 0.103 & $10^{10} \mathrm{M}{\odot} \mathrm{kpc}^{-2}$ \\
$x_1$ & 0.490 & - & kpc & &  $R_d$ & 4.750 & - & kpc \\
$y_1$ & 0.392 & - & kpc & &  $z_d$ & 0.151 & - & kpc \\
$z_1$ & 0.229 & - & kpc & &  $R_{cut}$ & 4.690 & - & kpc \\
$c_{||}$ & 1.991 & - & -- & &  $n_d$ & 1.540 & - & -- \\
$c_{\bot}$ & 2.232 & - & -- & &  $m_d$ & 0.716 & - & -- \\
$m$ & 0.975 & 0.873 & -- & &   &  &  &  \\
$\alpha$ & 0.626 & - & -- &  &  &  &  &  \\
$n$ & 1.940 & - & -- &  &  &  &  &  \\
$c$ & 1.342 & - & -- &  &  &  &  &  \\
$x_c$ & 0.751 & - & kpc &  &  &  &  &  \\
$y_c$ & 0.469 & - & kpc &  &  &  &  &  \\
$r_{cut}$ & 4.370 & - & kpc &  &  &  &  &  \\
\\
\multicolumn{4}{c}{\textbf{\underline{Barred component 2}}} & & \multicolumn{4}{c}{\textbf{\underline{Barred component 3}}} \\
\\
$\rho_2$ & 0.028 & 0.050 & $10^{10} \mathrm{M}{\odot} \mathrm{kpc}^{-3}$ & & $\rho_3$ & 1743.049 & - & $10^{10} \mathrm{M}{\odot} \mathrm{kpc}^{-3}$ \\
$x_2$ & 9.960 & 5.360 & kpc & & $x_3$ & 0.478 & - & kpc \\
$y_2$ & 0.713 & 0.959 & kpc & & $y_3$ & 0.267 & - & kpc \\
$z_2$ & 0.472 & 0.611 & kpc & & $z_3$ & 0.252 & - & kpc \\
$n_2$ & 3.050 & - & -- & & $n_3$ & 0.980 & - & -- \\
$c_{L,2}$ & 0.970 & - & -- & & $c_{L,3}$ & 1.880 & - & -- \\
$R_{2,out}$ & 2.840 & 3.190 & kpc & & $R_{3,out}$ & 2.204 & - & kpc \\
$R_{2,in}$ & 1.260 & 0.558 & kpc & & $R_{3,in}$ & 7.607 & - & kpc \\
$n_{2,out}$ & 16.200 & 16.731 & -- & & $n_{3,out}$ & -27.291 & - & -- \\
$n_{2,in}$ & 7.160 & 3.196 & -- & & $n_{3,in}$ & 1.630 & - & -- \\
\hline
\label{table 2}
\end{tabular}
\tablefoot{Values are compared to the analytical model by \protect\cite{sormani2022stellar}. S22 values are shown only when they differ from this work.\\
}
\end{table*}

We also compare our model with the hydrodynamical simulation presented in \cite{ridley2017nuclear}, which simulate the dynamics of central molecular zone, and is based on observations of NH$_{3}$ (1,1) emission from the HOPS survey and an improved potential from \cite{McMillan2017}. The R17 model includes 3-dimensional profiles for three components: bar, bulge, disc and halo. For the purpose of comparing the R17 model to our model's stellar profile in the Galactic centre alone, we exclude the halo component of R17, and confine our analysis to the inner 3\,kpc, where that component is negligible. 

\begin{figure*}
    \centering
    \includegraphics[width=\columnwidth]{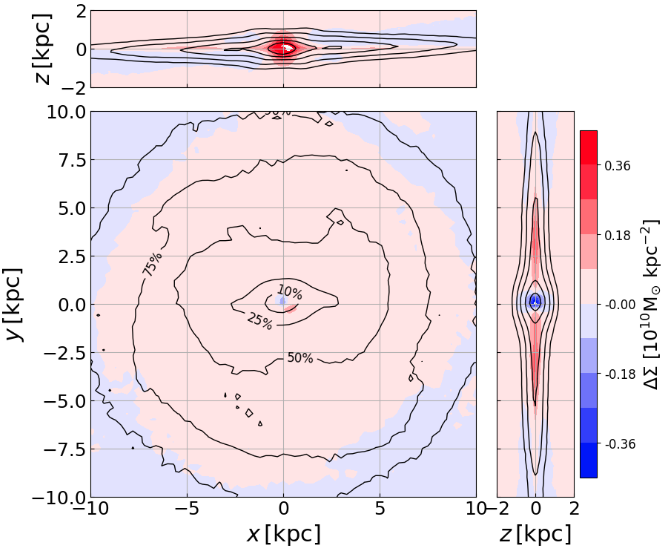}
    \includegraphics[width=\columnwidth]{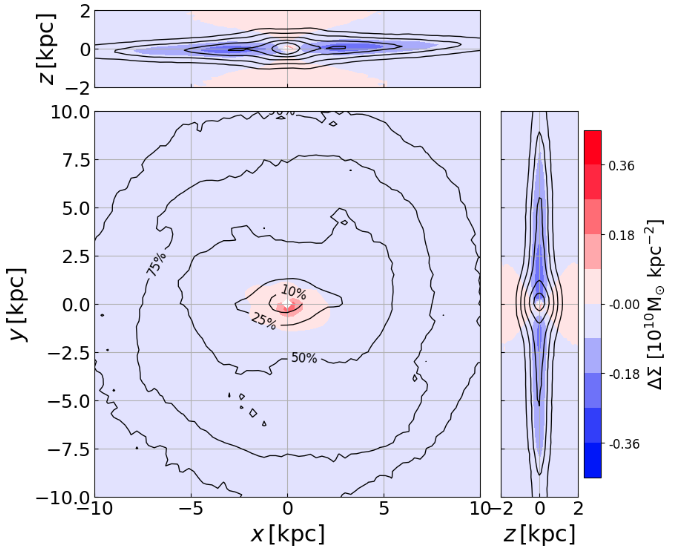}
    \caption{Residual column density maps of the stellar disc distribution within our model compared to two analytical models. The left panels show the comparison with the S22 analytical model, while the right panels compare with the analytical potential from R17. In both cases, the residuals are presented in the $xy$, $xz$, and $yz$ planes. Overlaid on the maps are contour plots representing cumulative fractions of the total stellar mass our model: $10\%, 25\%, 50\%, 75\%,$ and $90\%$.}
    \label{fig:residual_comparison2}
\end{figure*}

Figure~\ref{fig:1d R17} shows the 1D surface density profiles of R17 compared to our work and S22, in that inner region. Overall, this figure echoes the results from the S22 comparison: the main discrepancies appear in the $z$-direction, where the R17 profile is even less steep than S22, suggesting a deficit in mass in our models in the vertical direction, which increases with distance from the mid-plane. This can be better visualised in the 2D residual maps between the projected stellar surface densities of our model and those of the S22 and R17 models (Fig.\,\ref{fig:residual_comparison2}), where red/blue colours indicate a lack/excess of mass in our model, respectively. These plots reinforce the idea from the 1D analysis, that most of the discrepancy between our model and the two analytic descriptors (within the inner $\sim\,3$\,kpc) is on the $z$-direction.

The above analysis focuses on the projected surface densities (in 1 and 2D), hinting at a potential lack of mass in our model in the inner regions, but in order to assess how significant this missing mass might be, and how it might affect the inner kinematics, we also investigated the differences in terms of 3D enclosed mass between our model, and S22. In Section\,\ref{sec: kinematics}, we presented the comparison of the spherically enclosed mass, but given that most of the disparities between our model and the analytic profiles were in the $z$-direction, here we include a comparison of the enclosed mass within different $z$ slices. Note that for this purpose we used the S22 profile, but the conclusions would be near identical if considering the R17 profile, given that they follow each other closely up to $R<2.5$\,kpc and $|z|<1$\,kpc.

\begin{figure}
	\includegraphics[width=\columnwidth]{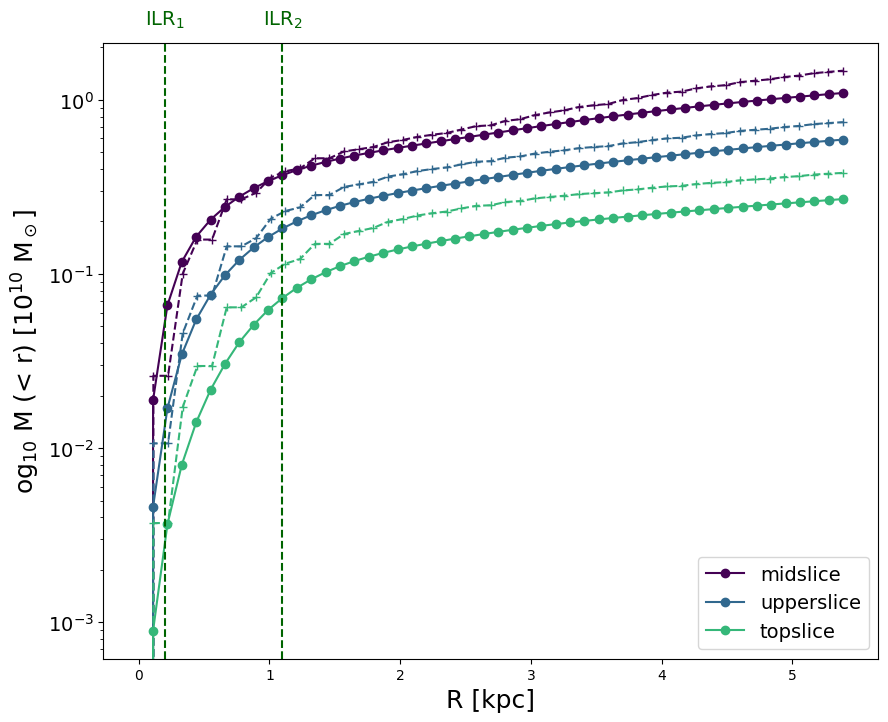}
    \caption{Cylindrical enclosed mass as a function of galactocentric radius for the analytical model from \protect\citep{sormani2022stellar} (cross-dashed lines), and our model (dot-solid lines). The different colours refer to the different $z$-slices: purple for the mid-slice ($0 < |z| < 0.25$\,kpc), blue for the upper-slice ($0.25 < |z| < 0.5$\,kpc) and green for the top-slice ($0.5 < |z| < 0.75$\,kpc). The position of the two ILRs in our model are shown as green vertical dashed lines.}
    \label{fig:enclosed mass cylindrical}
\end{figure}

Figure~\ref{fig:enclosed mass cylindrical} shows the enclosed mass as a function of the cylindrical radius for each slice, and we can see that in the mid-plane slice (i.e. mid-slice), the enclosed mass in our model closely matches the analytical model up to $R \sim2$~kpc, which is well past both ILRs. However, as we move outwards from the galactic mid-plane, our model presents an enclosed mass at the second ILR that is lower than the analytical model by 0.09 dex at intermediate $z$ (i.e. upper-slice) and 0.18 dex at higher $z$ (i.e. top-slice). This indicates that our model is more compacted in the mid-slice than the analytical model, with the mass concentration decreasing rapidly in the $z$-direction. We conclude, that although our model is indeed missing some mass in the inner galaxy, most of it is in the $z$-direction, and the differences in mass within the mid-plane of the galaxy are not significant enough to explain the lack of $x_2$-type of orbits. 

\section{The basics of Fourier series and application to our model}
\label{appendix: fourier}

An advantage of using numerical simulations for determining the number of spiral arms is that we have accurate positions for both the stellar and gaseous components, allowing us to apply techniques on the top-view column density distribution. This is otherwise challenging in observations of the Milky Way due to our inside view of the Galaxy. In order to have a better understanding of the spiral pattern present in our simulation, we use Fourier transformations, as studies of nearby galaxies \citep[e.g.][]{Fuchs1999} and numerical simulations \citep[e.g.][]{Bottema2003} have done in the past. We apply this technique to our model, both using gas and stars, to infer what spiral pattern emerges when using different tracers.
 
The Fourier series provides a powerful method for approximating or representing a periodic function through a combination of simple harmonic functions (sine and cosine). This approach is particularly useful for decomposing complex waveforms into basic components, enabling a detailed analysis of structures such as the spiral arms of galaxies \citep[e.g.][]{Davis2012}.

The Fourier series allow us to express a periodic function \(f(x)\) with a period of \(2L\) as a sum of sine and cosine functions. For a waveform \(f(x)\) and a period \(2L = \frac{2\pi}{k}\), where \(k = \frac{\pi}{L}\), the spatial frequency \(k\) relates to the wavelength \(L\) of the periodic function. Consequently, for any integer \(n\), the term \(nkx = \frac{n\pi x}{L}\) specifies the harmonics of the function. 

In this context, the Fourier coefficients — \(a_0\) (the constant term), \(A_n\) (the cosine coefficients), and \(B_n\) (the sine coefficients) were determined through integration over the function's period, as follows:

\begin{itemize}
    \item The constant term \(a_0\) is calculated as
    \[ a_0 = \frac{1}{L} \int_{-L}^{L} f(x) \, dx. \]
    This term represents the average or mean value of the function over one period.\\
    
    \item The cosine coefficients \(A_n\) for \(n = 1, 2, 3, \ldots\) are given by
    \[ A_n = \frac{1}{L} \int_{-L}^{L} f(x) \cos\left(\frac{n\pi x}{L}\right) \, dx. \]
    These coefficients measure the amplitude of the cosine components of the waveform, corresponding to the function's `even' symmetry parts.\\
    
    \item The sine coefficients \(B_n\) for \(n = 1, 2, 3, \ldots\) are determined as
    \[ B_n = \frac{1}{L} \int_{-L}^{L} f(x) \sin\left(\frac{n\pi x}{L}\right) \, dx. \]
    These coefficients capture the amplitude of the sine components of the waveform, relating to the function's `odd' symmetry parts.
\end{itemize}

The final form of $f(x)$ in terms of the Fourier coefficients $a_0$, $A_n$, and $B_n$ is given by the Fourier series expansion. This expansion expresses the periodic function $f(x)$ as a sum of its sinusoidal components:

\begin{equation}
    f(x) = \frac{a_0}{2} + \sum_{n=1}^{\infty} \left[A_n \cos\left(\frac{n\pi x}{L}\right) + B_n \sin\left(\frac{n\pi x}{L}\right)\right].
\end{equation}

We then proceed to identifying if there is a dominant periodicity mode in the sharpened images as a function of galactocentric distance via these Fourier coefficients. In our application, $f(x)$ represents the top-down surface density distribution of gas or stars within the simulation, where $x$ corresponds to the azimuthal angle in the plane of the galaxy, and $L$ relates to the radial extent of the region being analysed. The integer $n$ denotes the harmonic number and the resulting Fourier coefficients, $A_n$ and $B_n$, quantify the contribution of each harmonic $n$ to the overall Fourier coefficient at a given radius. The combined Fourier amplitude for each mode is given by

\begin{equation}
    C_n = \sqrt{A_n^2 + B_n^2}.
\end{equation}

This amplitude $C_n$ indicates the strength of each mode. In essence, each radial bin's density profile is approximated as a sum of these sinusoidal functions, with the coefficients indicating the presence and intensity of periodic patterns (e.g. spiral arms in our case). The process is repeated across multiple radial bins, and both sine and cosine coefficients (A$_{\mathrm{n}}$ and B$_{\mathrm{n}}$) are calculated via numerical integration using the trapezium approximation. The integration is performed within defined radial bins of $0.6$\,kpc separation, within a $15$ kpc radius. In Figure~\ref{fig:number spirals}, we plot the resulting $C_n$ amplitude as a function of galactocentric radius and harmonic number $n$. For our analysis, we look at whether there is a clear dominant mode $n$ for periodicity at any given radial range, as that could be indicative of a preferred number ($n$) of spiral arms present in the simulation.

\section{Spiral ridges}
\label{appendix: spiral ridges}

\begin{figure*}
    \includegraphics[width=\textwidth]{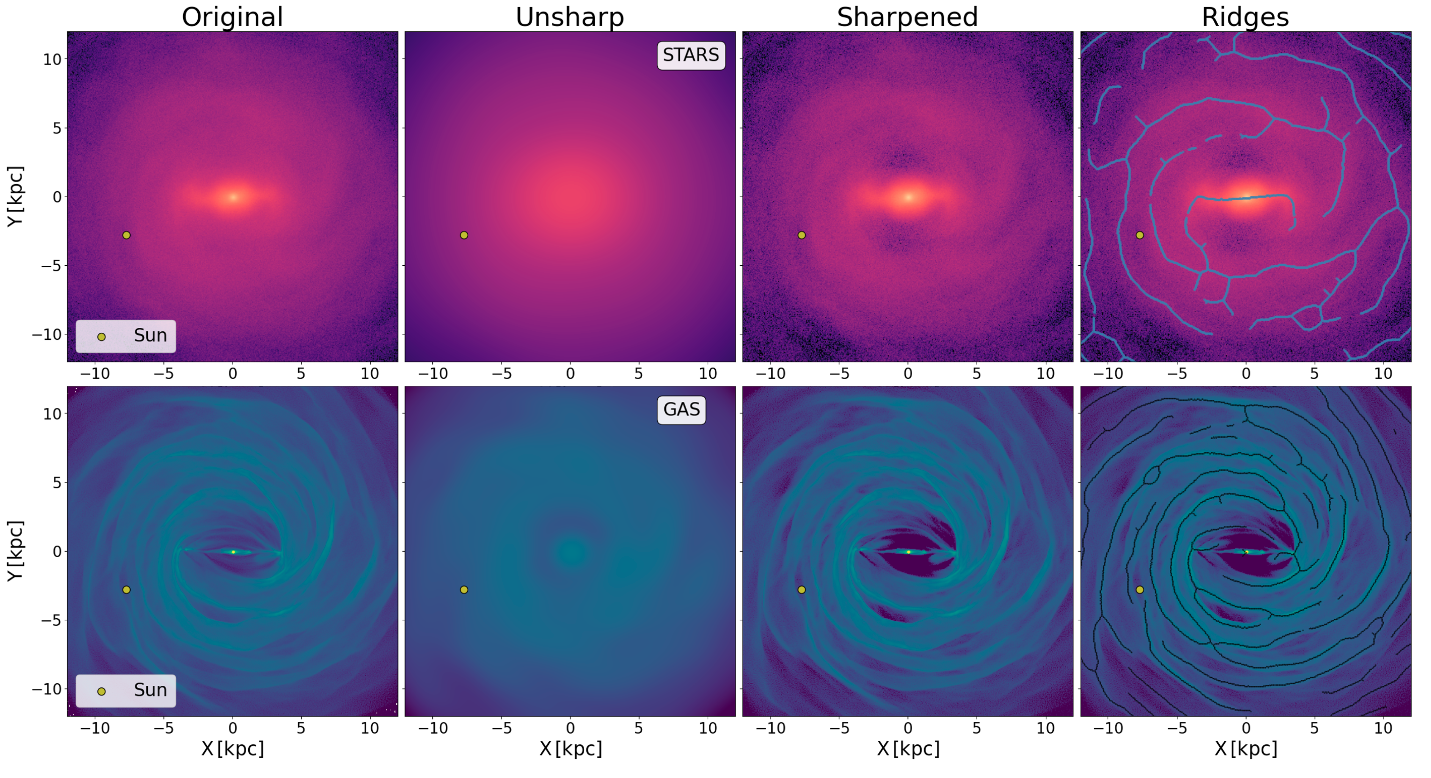}
    \caption{Maps illustrating the different steps involved in the unsharpening technique that we use to enhance the contrast of the spiral features from our model. From left to right: original, unsharp/smoothed, sharpened and spiral arm ridges images of the stellar (top) and gaseous (bottom) surface densities of our model. The galactic bar is positioned at the $y=0$ axis and the Sun's position is marked with a yellow dot.}
    \label{fig:Unsharp}
\end{figure*}

In order to investigate the link between the spiral arms and the kinematic pattern of our simulation, we extract the positions (or ``ridges'') of the spiral arms. To do so, we use the top-down surface density images of the stars and gas, after applying an unsharp masking technique to enhance the visibility of high surface density features. This step is particularly needed for the stellar map, given their low-level contrast of spiral features in the original projection. This process involves taking the original image ($I_{\rm{orig}}$) and create a background (unsharp) image ($I_{\rm{bg}}$) that captures only the vary large-scale fluctuations. The final sharpened image ($I_{\rm{final}}$) is then obtained by doing $I_{\rm{final}} = I_{\rm{orig}} + a(I_{\rm{orig}} - I_{\rm{bg}})$, where $a$ is the factor that regulates the level of sharpening. In our case, we take $a = 0.5$, and we produce $I_{\rm{bg}}$ by smoothing the original surface density maps with a Gaussian kernel of $2$\,kpc. For the gas, however, because the central $\sim 0.15$ kpc radial region has much higher surface densities than the background, applying this kernel would overestimate the overall background in the central region. To avoid this, we first mask out this central high density region and then apply the $2$\,kpc Gaussian kernel to the masked image, to produce the final unsharp map. The original, background and sharpened images from this procedure are shown in Fig.~\ref{fig:Unsharp} (columns 1-3).

To extract the ridges of these sharpened images, we follow the technique described in \cite{DuranCamacho2024}, where we extracted the 'skeletons' or footprints of the spiral arms from all \textit{l-v} maps in our range of simulations. We provide here a brief description of this technique, but we refer the reader to the original work for further details on the methodology. 

We start by slightly smoothing the sharpened images, by convolving with a small Gaussian smoothing kernel to ensure clarity and continuity in the arm structures. The kernel size was chosen based on the physical scale resolution we aim to investigate (in our case roughly of the order of a spiral arm width), and therefore we used $\sigma = 6$ pixels, which corresponds to 360\,pc (i.e. FWHM of $\sim 850$\,pc). We then compute the Hessian matrix of the image, with the respective eigenvalues, and create a binary mask of the image which traces the ridges of the map, by requiring the first eigenvalue to be negative (such that the surface is convex), with a stronger curvature than the second eigenvalue (to ensure that it is filamentary), and with an absolute value larger than a given threshold (to filter out the fainter structures). We then apply a binary erosion of $2$ pixels to this mask, in order to thinner the wider areas, and finally compute the medial axis of those eroded binary maps. The right-most column of Figure~\ref{fig:Unsharp} shows the extracted stellar (top) and gaseous (bottom) ridges overplotted on the corresponding sharpened images.

\section{Heliocentric kinematic maps}
\label{appendix: heliocentric kinematics}

In Section\,\ref{subsec: outer disc}, we compare the stellar kinematics of our model with those observed with {\it Gaia}, as a means to infer whether our model sees the same type of kinematical patterns, and understand how much of that can inform us on the underlying stellar and gaseous spiral pattern. Here, we provide more details on how we produced the heliocentric maps from the model, and we explore how the results may vary as a function of time, and specific viewing angle.

\begin{figure}
    \includegraphics[width=\linewidth]{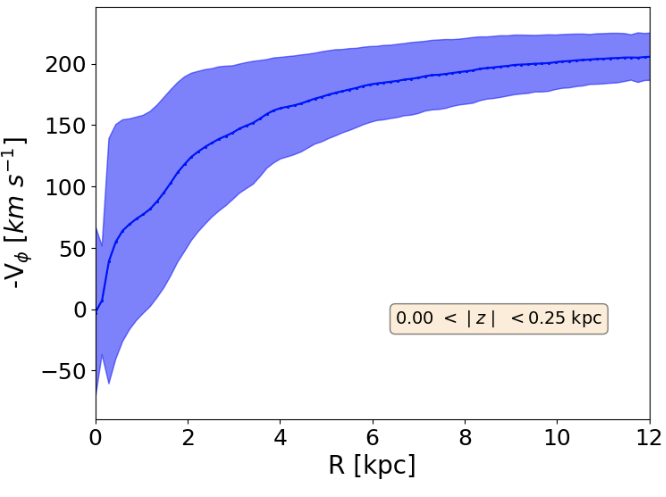}
    \caption{Median inverted tangential velocity (\(-V_\phi\)) as a function of radial distance from the Galactic centre for the combined stellar and gaseous components within the mid-slice of our simulation. The shaded area corresponds to the standard deviation from the mean velocity value of all particles included within that radial bin.}
    \label{fig:velocity_analysis}
\end{figure}

In a similar way to the observations by \citet{khanna2023measuring}, we obtain the velocity residuals by contrasting the actual velocity fields observed in our simulation against an idealised rotation curve. We compute the idealised rotation curve by assuming perfect circular rotation, with no radial ($V_R$) or vertical ($V_z$) velocities. We then consider the tangential velocity ($-V_{\phi,\mathrm{ideal}}$) as being the median velocity of all particles (stellar and gaseous) included in each radial bin. This idealised rotation curve can be seen in Fig.~\ref{fig:velocity_analysis}. We then obtain the velocity residual, $\Delta(-V_\phi)$, by subtracting this idealised $-V_{\phi,\mathrm{ideal}}$ from the actual $-V_\phi$ observed for the particles within that segment. These residuals will identify non-circular motions and perturbations within the model. We focus our comparison on the trends seen on mid-plane of the galaxy (i.e. $0 < z < 0.25$\,kpc)\footnote{On their work, \cite{khanna2023measuring} explore higher $z$ slices up to $z=0.75$\,kpc. However, velocity trends were similar across the different $z$-regions, with less extreme motions as moving away from the Galactic plane. When looking at different heights within our model, results are near identical for all values of $z$.}.

\begin{figure}
    \includegraphics[width=0.48\textwidth]{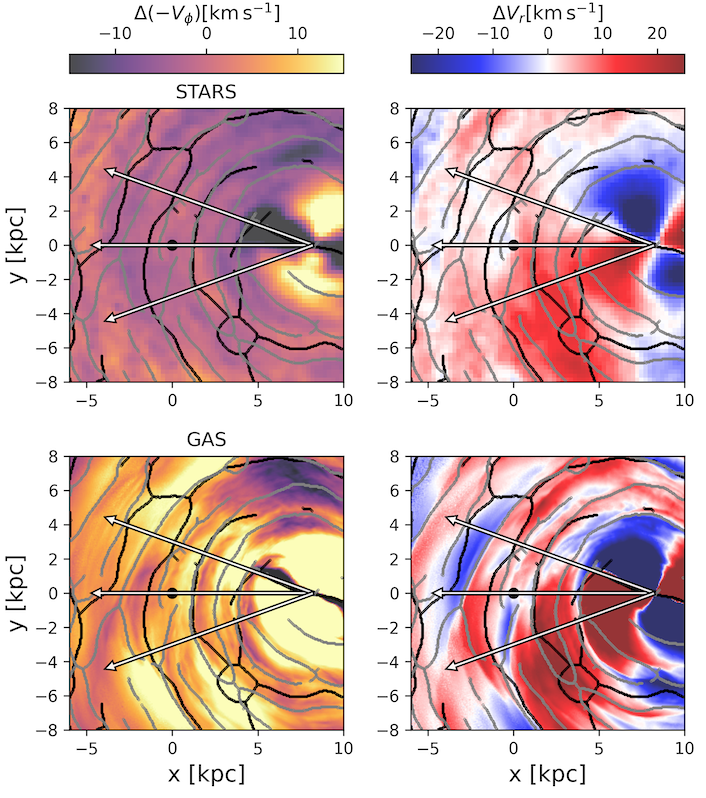}
    \caption{Heliocentric velocity residual maps for $\Delta(-V_\phi)$ (left) and $\Delta V_R$ (right) for the stars (top) and gas (bottom) in the mid-slice of our model ($ |z| < 0.25$\,kpc). The ridges of the spiral arm segments extracted from the sharpened images of Fig.~\ref{fig:Unsharp}, are superimposed for comparison - grey for gaseous and black for stellar ridges. The Galactic centre is at (8.275,0), and the Sun's position is marked with a black dot. The different cross-sections studied in Section~\ref{Subsec: gas disc kinematics} are represented with white arrows.}
    \vspace{-0.5cm}
    \label{fig: as drimmel}
\end{figure}

In Figure~\ref{fig:1d velocity res - gaia stars and gas}, we presented the heliocentric velocity residuals, \(\Delta(-V_\phi)\) and \(\Delta V_R\), for the data from the \textit{Gaia} DR3 study by \cite{khanna2023measuring} (top), alongside our model's. For that figure, we rotate our frame to match that used in the \textit{Gaia} work, such that the $x$ axis is the line connecting the observer and the Galactic centre, and the sun is at $(x,y)=(0,0)$, and we restricted our coverage and resolution of the stellar component to match that of \cite{khanna2023measuring} for a direct comparison. Here we include an extended version of our heliocentric velocity residual maps (Figure~\ref{fig: as drimmel}), so that we cover the Galactic centre \citep[and thus comparable to the \textit{Gaia} maps from][]{drimmel2023}. In Figure~\ref{fig: as drimmel}, we also overlay, as arrows, the different cross-sections studied in Section~\ref{Subsec: gas disc kinematics} and Appendix\,\ref{appendix: cross-sections}.

Between the Sun and the Galactic centre, the observations find predominantly radially outward motions, and our model finds a similar trend for the stars inwards of the observer position. The observations also find that around the Local arm, there is a change of sign, and the stellar motions become predominantly inward motions for larger galactocentric radii. Again, we find a similar trend in our model, albeit less sharp in amplitude. Intuitively, we would expect the convergence between the inward and outward motions to correspond to the position of the stellar spiral arm, but from our models, there is not always an obviously correlation between the velocity pattern of the stars, and the position of the stellar spiral arms.

\begin{figure*}
    \includegraphics[width=\textwidth]{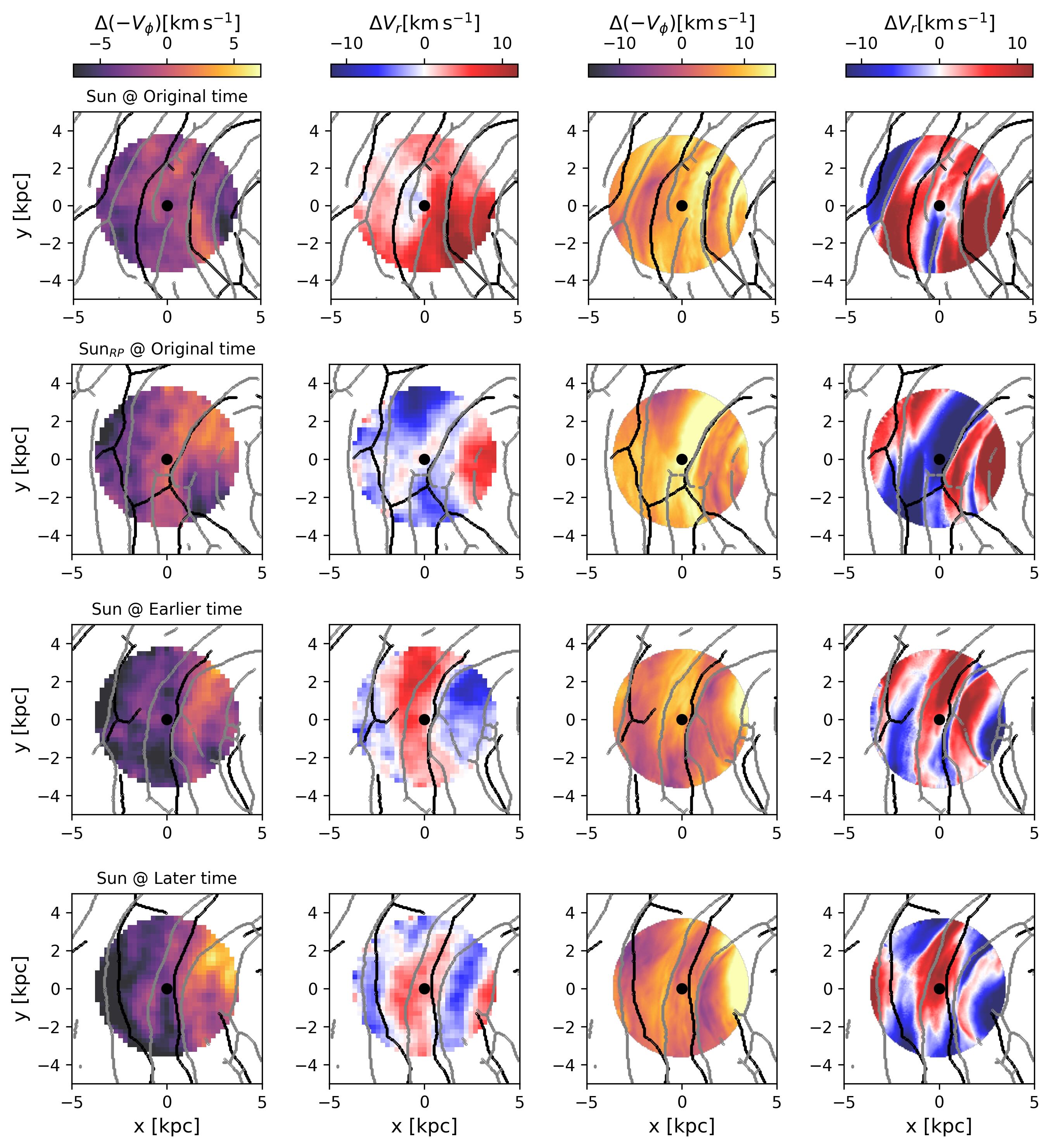}
    \caption{Comparison of the velocity residuals of our model: around the sun's original location at the original time (top row - as shown in Fig.\,\ref{fig:1d velocity res - gaia stars and gas}); at the sun's reflection point, Sun$_{\mathrm{RP}}$, at the original time (second row); around the sun's location, 10\,Myrs earlier than the original time (third row); and finally, around the sun's location, 10\,Myrs later than the original time (bottom row). The first two columns present the tangential and radial velocity residuals for the stars, while the two last columns show those of the gas. The ridges of the spiral arm segments, extracted from the respective sharpened images, are overlaid in black for the stellar ridges and grey for the gas ridges.}
    \label{fig:1d velocity res - gas stars inverted before and after}
\end{figure*}

Finally, in order to investigate how a different viewing angle and different times can alter the results from Section\,\ref{sec: kinematics}, in Fig.\,\ref{fig:1d velocity res - gas stars inverted before and after} we show the residual velocity maps around the solar neighbourhood for our model at two different times (10\,Myr before and after the optimal time shown in Fig.\,\ref{fig:1d velocity res - gaia stars and gas}), as well as for the Sun reflection point at the optimal time, Sun$_{\rm{RP}}$ (i.e. located at 180$^{\circ}$ rotated around the Galactic centre, such that the angle between the bar and the Sun$_{\rm{RP}}$ remains at 20$^{\circ}$). This figure shows that the kinematical patterns (in particular changes of sign, or positions of maxima and minima) sometimes line up with the ridges of the stellar spiral arms and/or the gaseous ones, but there is no systematic trend. More interestingly, we can also see that these patterns seem to change quite rapidly: for instance, in a space of 20\,Myr, the pattern of the radial velocities reverse such that, as we move from the outer galaxy through to the Galactic centre, it goes from outflow-inflow (red-to-blue) 10\,Myrs before, to inflow-outflow (blue-to-red) at the original optimal time, to outflow-inflow again (red-to-blue) 10\,Myrs after. The dynamic nature of these patterns is also nicely captured in a more detailed video of the time evolution of the system, available on the FFOGG website\footnote{https://ffogg.github.io}.

Overall, this exercise shows that our model - at the time and location as originally selected in \citet{DuranCamacho2024} - successfully reproduces the kinematic imprints observed in the latest \textit{Gaia} results, but it also underscores the rapid evolution of the kinematical pattern over time, and the difficulty in associating any given kinematical signature to the underlying presence of a spiral arm.

\section{Stellar and gaseous cross-sections}
\label{appendix: cross-sections}

\begin{figure*}
        \includegraphics[width=\textwidth]{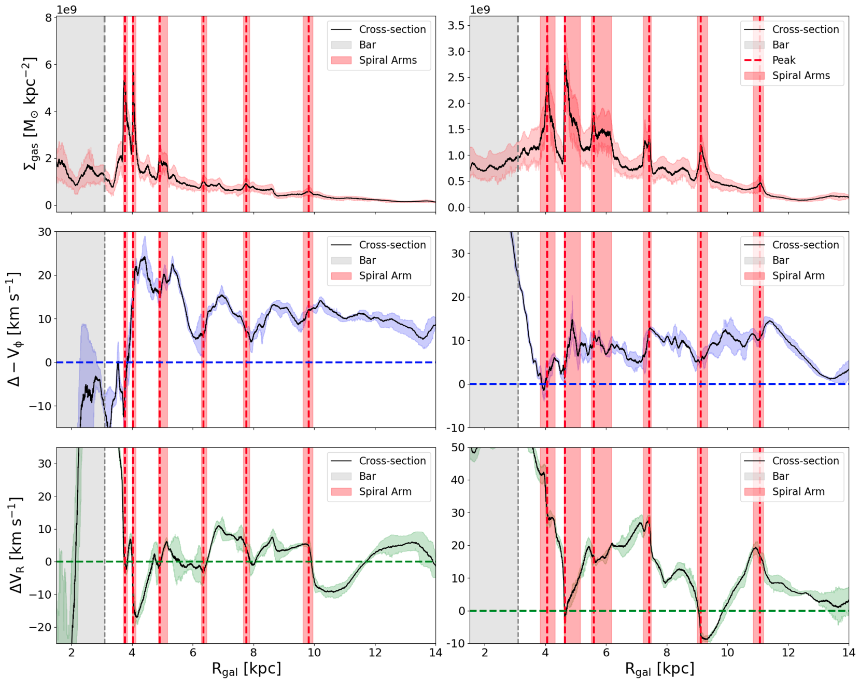}
    \caption{Same as Fig.~\ref{fig:LOS_0deg}, but for a line of sight of $\pm 20^{\circ}$ (left, right respectively) from the Sun's position with origin at the Galactic centre. The galactic bar is aligned with the line of sight of left column.}
    \label{fig:LOS_20deg}
\end{figure*}

We present in Fig.~\ref{fig:LOS_20deg} the analogous cross-sections to Fig.~\ref{fig:LOS_0deg} positioned $\pm 20^{\circ}$ above (left) and below (right) the line connecting the Galactic centre to the Sun. Similarly to the results shown in Section~\ref{Subsec: gas disc kinematics}, there are a number of gaseous spiral segments identified along both directions. The left column, corresponding to the $+ 20^{\circ}$ direction, aligns with the major axis of the galactic bar, which explains why the tangential residuals in the bar region exhibit a positive slope starting from negative values, in contrast with the other two cases. This region intersects the strong negative feature seen along the major axis of the bar in the tangential residuals, as shown in Fig.\,\ref{fig: as drimmel} in Appendix\,\ref{appendix: heliocentric kinematics}. Absolute changes within the bar region are of the order of $\sim 70$\,km\,s$^{-1}$, while in the other direction they are $\sim 30$\,km\,s$^{-1}$ across a radius of $\sim 3$\,kpc. Beyond this point, similar trends are observed in both cross-sections. Absolute changes in tangential and radial residuals are larger closer to the centre, where the gravitational potential is stronger and affected by the bar. Changes in tangential velocities  across spiral arms typically correspond to a velocity gradient of $\sim 10$\,km\,s$^{-1}$\,kpc$^{-1}$, with gas generally decelerated within the spiral arm compared to the inter-arm region. Regarding radial residuals, converging points (downward slopes) occur within the spiral arms, while diverging points (upward slopes) are found in inter-arm regions. A notable example of diverging point can be seen in the bottom panel of the right column in Fig.\ref{fig:LOS_20deg} at a distance of $\sim 10$\,kpc. This area appears to be between two spiral arms and the gas in the inter-arm region diverges towards each arm, with a velocity gradient of $\sim 15$\,km\,s$^{-1}$\,kpc$^{-1}$. 

\section{Spiral arm evolution across time}
\label{appendix: arm evolution}

\begin{figure*}
        \includegraphics[width=\textwidth]{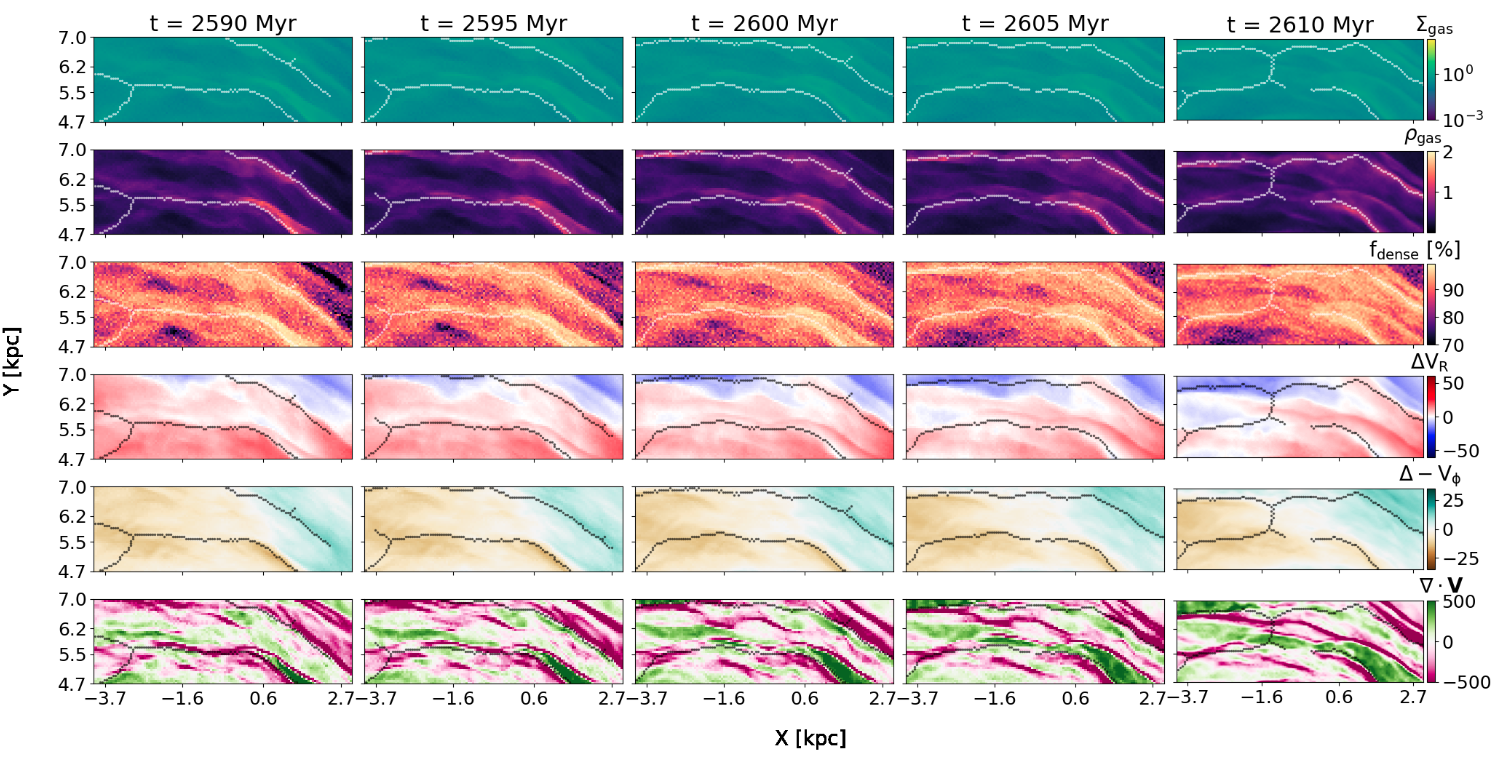}
    \caption{Time evolution of the spiral ridge analysed in Section~\ref{Subsec:time_evolution_spiral_arm} as a function of distance a long the ridge, shown for multiple time snapshots (columns). Each row corresponds to a different diagnostic map, from top to bottom: surface density, volume density, radial and tangential velocity residuals and velocity divergence. Coloured scatter points represent spiral ridge values obtained as explained in Appendix~\ref{appendix: spiral ridges}.}
    \label{fig:arm_times}
\end{figure*}

We examine the connection between spiral arm formation and dissolution and the large-scale kinematics in Sect.\ref{Subsec:time_evolution_spiral_arm} by following a representative gaseous arm segment over time. The segment is selected from the top-down surface-density map at $t=2.6$\,Gyr (Fig.\ref{fig: zoom_region}, inset box). Its subsequent evolution is shown in Fig.\ref{fig:arm_times} over a 20\,Myr interval, sampled every 5\,Myr (each column). From top to bottom, Fig.\ref{fig:arm_times} shows the surface density, volume density, fraction of dense gas, radial and tangential velocity residuals, and the velocity divergence. We then use these maps to characterise the evolution of these properties along the spiral arm ridge as a function of time in Fig.~\ref{fig:arm_evolution_along_ridge}.

In the first column of Fig.~\ref{fig:arm_times}, we identify an arm segment, hereafter referred to as \textit{Section 1}, located near \( x \sim -2.5\,\mathrm{kpc} \), where the spiral ridge is relatively faint. At this time, the divergence map (bottom row) shows only mild convergence in this segment, i.e. $\nabla\cdot\mathbf{V}<0$ (light pink). Over time, the same location becomes more strongly convergent and the ridge intensifies, reaching visibly higher surface and volume densities in the later snapshots $\sim 15-20$~Myr later. This is consistent with the evolution of the radial velocities (fourth row): at late times, \textit{Section 1} (centred approximately at R$\simeq -1.6$ kpc) lies at the interface between outflowing ($\Delta V_R>0$, red) and inflowing ($\Delta V_R<0$, blue) gas, i.e. a convergence zone that promotes ridge growth.

On the other side of this spiral arm section, we identified another region (\textit{Section 2}), located near \( x \sim 0\,\mathrm{kpc} \), which breaks/dissolves over this 20\,Myr period. This ridge starts at high density with highly converging motions, but as time evolves, it becomes increasingly divergent and breaks apart. The bottom panels of this figure show that adjacent to the location of this spiral arm ridge, there is a highly divergent zone (green region initially located at R$\sim 0.8$ kpc) that develops and grows with time, effectively `splitting' and spreading this spiral arm segment into a wider and lower density feature. The location of the arm in the first timestep is pushed downwards (following the lower edge of the split), while the upper edge is pushed outwards (and will eventually merge with the other visible spiral arm in the figure). This `opening' of the spiral arms is not just characteristic of this particular segment, but a recurring feature of the spiral arms in the simulation - best seen in the galaxy-wide time-evolution video available on the FFOGG webpage.

\end{appendix}
\end{document}